\documentclass[sts]{imsart}
\RequirePackage[OT1]{fontenc}
\usepackage{amsthm,amsmath,natbib}
\RequirePackage[colorlinks,citecolor=blue,urlcolor=blue]{hyperref}

\startlocaldefs
\numberwithin{equation}{section}
\theoremstyle{plain}

\endlocaldefs

\usepackage{graphicx} 
\usepackage{caption}
\usepackage[margin=20pt]{subcaption}
\usepackage{booktabs}
\usepackage{mathtools}
\usepackage{adjustbox}
\usepackage[flushleft]{threeparttable}
\usepackage{bm}
\usepackage{amssymb}

\usepackage{enumitem}

\usepackage{multirow}

\usepackage{comment}
\theoremstyle{definition}

\newtheorem*{theorem*}{Theorem}
\newtheorem*{rmk*}{Remark}

\newtheorem{example}{Example}

\newtheorem*{corollary*}{Corollary}

\def\Tau{\bm{\mathcal{T}}}

\usepackage{color}

\usepackage{xcolor}

\usepackage[sc]{mathpazo}

\usepackage[textsize=scriptsize, shadow, color = yellow, textwidth = 3.6cm]{todonotes}

\begin{document}

\begin{frontmatter}
\title{Bayesian Analysis of Rank Data with Covariates and Heterogeneous Rankers
}
\runtitle{Bayesian Analysis of Rank Data}

\begin{aug}

\author{\fnms{Xinran} \snm{Li},
     \ead[label=e2]{xinranli@illinois.edu}}
\author{\fnms{Dingdong} \snm{Yi}
	\ead[label=e1]{yidingdong@gmail.com}}
	\and
\author{\fnms{Jun S.} \snm{Liu}
\ead[label=e3]{jliu@stat.harvard.edu}}

\runauthor{ X. Li, D. Yi, and J. S. Liu}

\affiliation{University of Illinois at Urbana-Champaign and Harvard University}

\address{\footnotesize
    Xinran Li, Department of Statistics, University of Illinois, Champaign, IL 61820 \printead{e2}.
	Dingdong Yi, Department of Statistics, Harvard University, Cambridge, MA 02138 \printead{e1}.
    Jun S. Liu, Department of Statistics, Harvard University, Cambridge, MA 02138 \printead{e3}.}
\end{aug}

\begin{abstract}
Data in the form of ranking lists are frequently encountered, and combining ranking results from different sources can  potentially generate a better ranking list and help understand  behaviors of the rankers.  Of interest here are the rank data 
under the following settings: 
(i) covariate information available for the ranked entities; (ii) rankers of varying qualities or 
having different opinions; and (iii) incomplete ranking lists for non-overlapping subgroups. We review some key ideas  built around  the Thurstone model family by researchers in the past few decades 
and provide a unifying approach for  Bayesian Analysis of Rank data with Covariates (BARC) and its extensions in handling heterogeneous rankers.  
With this Bayesian framework, we can
study rankers' varying quality, cluster   
rankers' heterogeneous opinions, and measure the corresponding uncertainties. 
To enable an efficient Bayesian inference, we advocate a parameter-expanded Gibbs sampler to sample from the target posterior distribution.
The posterior samples also result in a Bayesian aggregated ranking list, with credible intervals quantifying its uncertainty. 
We investigate and compare performances of the proposed methods and other rank aggregation methods in both simulation studies and two real-data examples. 
\end{abstract}


\end{frontmatter}

\section{Introduction}\label{sec:intro}
\subsection{Motivating examples}
Rank data are rather prevailing these days, and combining ranking results from different sources is a common problem. 
Well-known rank aggregation problems range from the election problem back in the 18th century \citep{borda1781memoire} to search engine results aggregation in modern days \citep{dwork2001rank, liu2009learning}. In many cases, there are variations and complications associated with rank data. Sometimes, there are relevant covariates of the ranked entities while the ranking lists are highly incomplete. Also, the rankers are likely heterogeneous. Here, we illustrate the problem in detail using the following two examples. 

\begin{example}[NFL Quarterback Ranking]\label{eg:nfl}
	During the National Football League (NFL) season, experts from different websites, such as \url{espn.com} and \url{nfl.com}, provide weekly ranking lists of players by position. 
	For example, 
	Table \ref{tab:NFL_rank_data} shows the ranking lists of the NFL starting quarterbacks from 13 experts in week 12 of season 2014. The ranking lists can help football fans better predict the performance of the quarterbacks in the coming week and even place bets in online fantasy sports games. After collecting ranking lists from the experts, most websites aggregate them using arithmetic means. 
	Besides rankings, some summary statistics of the NFL players are also available online. 
	For example, 	Table \ref{tab:NFL_covariate} shows the statistics of
	the ranked quarterbacks prior to week 12 of season 2014. 
	Not surprisingly, in addition to watching football games, the experts may also use these summary statistics when ranking quarterbacks. 
	\begin{table}[htb]
		\scriptsize
		\caption{Ranking lists of NFL starting quarterbacks from 13 different experts, as of week 12 in the 2014 season.
		The first column shows the players' names, and the remaining columns show the ranked positions of these players from the 13 experts. 
		}
		\label{tab:NFL_rank_data}
		\centering
		\begin{tabular}{crrrrrrrrrrrrr}
			\toprule
			Player & $\tau_1$ & $\tau_2$ & $\tau_3$ & $\tau_4$ & $\tau_5$ & $\tau_6$ & $\tau_7$ & $\tau_8$ & $\tau_9$ & $\tau_{10}$ & $\tau_{11}$ & $\tau_{12}$ & $\tau_{13}$  \\ 
			\midrule
			Andrew	Luck	&	1	&	1	&	1	&	3	&	3	&	1	&	1	&	1	&	1	&	1	&	1	&	1	&	1	\\
			Aaron	Rodgers	&	2	&	3	&	4	&	2	&	1	&	2	&	3	&	3	&	2	&	2	&	3	&	4	&	3	\\
			Peyton	Manning	&	3	&	2	&	5	&	4	&	2	&	3	&	2	&	2	&	3	&	4	&	4	&	2	&	2	\\
			Tom	Brady	&	4	&	7	&	3	&	5	&	4	&	5	&	4	&	6	&	4	&	3	&	6	&	8	&	4	\\
			Tony	Romo	&	9	&	5	&	6	&	1	&	5	&	4	&	5	&	4	&	5	&	5	&	7	&	6	&	6	\\
			Drew	Brees	&	10	&	4	&	2	&	8	&	9	&	7	&	7	&	5	&	7	&	6	&	2	&	3	&	5	\\
			Ben	Roethlisberger	&	6	&	8	&	7	&	7	&	7	&	6	&	6	&	10	&	6	&	7	&	5	&	7	&	7	\\
			Ryan	Tannehill	&	5	&	6	&	13	&	6	&	11	&	8	&	8	&	7	&	9	&	9	&	8	&	5	&	8	\\
			Matthew	Stafford	&	8	&	9	&	11	&	13	&	8	&	9	&	9	&	8	&	8	&	8	&	9	&	9	&	9	\\
			Mark	Sanchez	&	22	&	10	&	9	&	9	&	16	&	10	&	10	&	9	&	10	&	10	&	12	&	12	&	12	\\
			Russell	Wilson	&	12	&	13	&	17	&	10	&	10	&	12	&	11	&	12	&	11	&	12	&	11	&	14	&	15	\\
			Philip	Rivers	&	7	&	14	&	15	&	20	&	6	&	17	&	17	&	11	&	16	&	15	&	14	&	10	&	10	\\
			Cam	Newton	&	18	&	12	&	8	&	17	&	19	&	11	&	14	&	14	&	14	&	16	&	21	&	13	&	14	\\
			Eli	Manning	&	17	&	--	&	18	&	19	&	14	&	19	&	12	&	13	&	12	&	13	&	16	&	23	&	11	\\
			Matt	Ryan	&	21	&	17	&	19	&	15	&	20	&	15	&	15	&	15	&	13	&	11	&	20	&	21	&	13	\\
			Andy	Dalton	&	15	&	--	&	14	&	--	&	17	&	14	&	16	&	20	&	15	&	14	&	19	&	22	&	16	\\
			Alex	Smith	&	16	&	11	&	21	&	16	&	18	&	18	&	18	&	16	&	20	&	21	&	13	&	11	&	17	\\
			Colin	Kaepernick	&	11	&	16	&	16	&	11	&	12	&	16	&	21	&	17	&	19	&	18	&	22	&	16	&	21	\\
			Joe	Flacco	&	24	&	15	&	12	&	14	&	24	&	13	&	13	&	18	&	18	&	20	&	15	&	15	&	19	\\
			Jay	Culter	&	13	&	18	&	10	&	12	&	13	&	21	&	19	&	19	&	17	&	17	&	23	&	20	&	18	\\
			Josh	McCown	&	14	&	19	&	22	&	18	&	15	&	22	&	22	&	21	&	21	&	19	&	18	&	17	&	23	\\
			Drew	Stanton	&	20	&	20	&	--	&	22	&	22	&	20	&	20	&	23	&	22	&	22	&	10	&	19	&	20	\\
			Teddy	Bridgewater	&	23	&	21	&	20	&	21	&	23	&	23	&	23	&	22	&	23	&	24	&	17	&	18	&	22	\\
			Brian	Hoyer	&	19	&	--	&	--	&	--	&	21	&	24	&	24	&	24	&	24	&	23	&	24	&	24	&	24	\\
			\bottomrule
		\end{tabular}%
		\\
		\begin{tablenotes}
		\item 
		Source: \url{fantasy.nfl.com/research/rankings}, \url{www.fantasypros.com/nfl/rankings/qb.php}.
		\end{tablenotes}
	\end{table}
	\begin{table}[htb]
		\scriptsize
		\caption{Relevant statistics of the ranked quarterbacks, prior to week 12 of the 2014 NFL season. From left to right, the statistics stand for: number of games played; pass completion percentage; passing attempts per game; average passing yards per attempt; touchdown percentage; intercept percentage; running attempts per game; running yards per attempt; running first down percentage.}
		\centering
		\label{tab:NFL_covariate}
        \resizebox{\columnwidth}{!}{%
			\begin{tabular}{lrrrrrrrrrrr}
				\toprule
				
				Player & G & Pct & Att & Avg & Yds & TD & Int & RAtt & RAvg & RYds & R1st \\ 
				\midrule
				Andrew Luck &  11 & 63.40 & 42.20 & 7.80 & 331.00 & 6.30 & 2.20 & 4.20 & 4.20 & 17.50 & 30.40 \\ 
				Aaron Rodgers &  11 & 66.70 & 31.10 & 8.60 & 268.80 & 8.80 & 0.90 & 2.50 & 6.40 & 16.20 & 50.00 \\ 
				Peyton Manning &  11 & 68.10 & 40.20 & 8.00 & 323.50 & 7.70 & 2.00 & 1.50 & -0.50 & -0.70 & 0.00 \\ 
				Tom Brady &  11 & 65.00 & 37.90 & 7.20 & 272.50 & 6.20 & 1.40 & 1.70 & 0.70 & 1.30 & 21.10 \\ 
				Tony Romo &  10 & 68.80 & 29.50 & 8.50 & 251.90 & 7.50 & 2.00 & 1.50 & 2.50 & 3.70 & 20.00 \\ 
				Drew Brees &  11 & 70.30 & 42.00 & 7.60 & 317.40 & 4.80 & 2.40 & 1.70 & 2.80 & 4.90 & 26.30 \\ 
				Ben Roethlisberger &  11 & 68.30 & 37.50 & 7.90 & 297.30 & 5.80 & 1.50 & 1.90 & 1.10 & 2.10 & 19.00 \\ 
				Ryan Tannehill &  11 & 66.10 & 35.40 & 6.60 & 234.70 & 5.10 & 2.10 & 3.70 & 6.70 & 25.10 & 36.60 \\ 
				Matthew Stafford &  11 & 58.80 & 37.70 & 7.10 & 267.50 & 3.10 & 2.40 & 2.80 & 2.00 & 5.60 & 16.10 \\ 
				Mark Sanchez &   4 & 62.30 & 36.50 & 8.10 & 296.80 & 4.80 & 4.10 & 3.50 & 0.60 & 2.00 & 7.10 \\ 
				Russell Wilson &  11 & 63.60 & 28.50 & 7.10 & 202.70 & 4.50 & 1.60 & 7.60 & 7.70 & 58.50 & 45.20 \\ 
				Philip Rivers &  11 & 68.30 & 33.00 & 7.80 & 257.70 & 6.10 & 2.50 & 2.50 & 2.50 & 6.40 & 25.00 \\ 
				Cam Newton &  10 & 58.60 & 33.30 & 7.20 & 239.20 & 3.60 & 3.00 & 6.40 & 4.60 & 29.30 & 37.50 \\ 
				Eli Manning &  11 & 62.30 & 36.90 & 7.00 & 257.50 & 5.20 & 3.00 & 0.80 & 3.80 & 3.10 & 33.30 \\ 
				Matt Ryan &  11 & 65.10 & 38.50 & 7.20 & 278.70 & 4.50 & 2.10 & 1.60 & 4.30 & 7.10 & 33.30 \\ 
				Andy Dalton &  11 & 62.40 & 30.70 & 7.10 & 219.40 & 3.60 & 3.00 & 3.80 & 2.50 & 9.50 & 33.30 \\ 
				Alex Smith &  11 & 65.10 & 29.70 & 6.80 & 201.00 & 4.00 & 1.20 & 3.20 & 5.50 & 17.40 & 25.70 \\ 
				Colin Kaepernick &  11 & 61.70 & 31.50 & 7.50 & 237.70 & 4.30 & 1.70 & 6.80 & 4.50 & 30.50 & 22.70 \\ 
				Joe Flacco &  11 & 63.20 & 34.10 & 7.40 & 251.30 & 4.80 & 2.10 & 2.00 & 1.70 & 3.40 & 45.50 \\ 
				Jay Cutler &  11 & 66.80 & 36.40 & 7.10 & 256.80 & 5.50 & 3.00 & 2.90 & 3.90 & 11.30 & 28.10 \\ 
				Josh McCown &   6 & 60.40 & 30.30 & 7.40 & 225.00 & 3.80 & 4.40 & 2.70 & 5.80 & 15.30 & 50.00 \\ 
				Drew Stanton &   6 & 53.60 & 25.20 & 7.10 & 178.20 & 3.30 & 2.00 & 3.00 & 2.00 & 6.00 & 22.20 \\ 
				Teddy Bridgewater &   8 & 60.30 & 32.80 & 6.40 & 211.10 & 2.30 & 2.70 & 3.50 & 4.60 & 16.10 & 32.10 \\ 
				Brian Hoyer &  11 & 55.90 & 33.20 & 7.80 & 260.40 & 3.00 & 2.20 & 1.80 & 0.90 & 1.50 & 20.00 \\ 
				\bottomrule
			\end{tabular}%
			}
		\begin{tablenotes}
			\item Source: \url{www.nfl.com/stats}.
		\end{tablenotes}
	\end{table}
\end{example}

In Example \ref{eg:nfl}, 
according to Table \ref{tab:NFL_rank_data}, most experts give very similar ranking lists, with a few exceptions such as experts 1 and 5. 
Besides, 
some rankers do not place the players in Table \ref{tab:NFL_rank_data} on their top 24 lists, making the ranking lists incomplete. 
Therefore, 
it is of interest to understand how the rankings may be dependent of the available summary statistics (i.e., covariates), whether the experts (i.e., rankers) are consistent in using these covariates when ranking the players, and whether they (the rankers) have different qualities or different opinions. 
Another goal is to obtain an aggregated ranking list of all players taking into account the covariate information of the players and the potential heterogeneity of the experts,  
which hopefully can improve the accuracy of rank aggregation compared to the simple arithmetic means. 


\begin{example}[Orthodontics treatment evaluation ranking]\label{eg:ortho} 
	In 2009, 69 orthodontics experts were invited by the School of Stomatology at Peking University to evaluate the post-treatment conditions of 108 medical cases \citep{song2015validation}. In order to make the evaluation easier for the experts, cases were divided into 9 groups, each containing 12 cases. For each group of the cases, each expert evaluated the conditions of all 12 cases and provided a within-group ranking list, mostly based on their personal experiences and judgments of the cases' teeth records.
	In the meantime, using each case's plaster model, cephalometric radiograph, and photograph, the School of Stomatology located key points, measured their distances and angles that are considered to be relevant features for diagnosis, and summarized these features in terms of peer assessment rating (PAR) index \citep{richmond1992development}. 
	Table \ref{tab:ortho_rank_data} shows 15 of the 69 ranking lists for two groups, and Table \ref{tab:ortho_covariate} shows the corresponding features for these two groups. 
	\begin{table}[htb]
		\scriptsize
		\centering
		\caption{Ranking lists for Groups A and H, two of the 9 groups in Example 2.
		The first column shows the groups and indices for the cases, and the remaining columns show the within-group ranked positions of these cases from 15 experts. 
		}
		\label{tab:ortho_rank_data}
			\begin{tabular}{rrrrrrrrrrrrrrrrrrrrr}
				\toprule
				& $\tau_{1}$ & $\tau_{2}$ & $\tau_{3}$ & $\tau_{4}$ & $\tau_{5}$ & $\tau_{6}$ & $\tau_{7}$ & $\tau_{8}$ & $\tau_{9}$ & $\tau_{10}$ & $\tau_{11}$ & $\tau_{12}$ & $\tau_{13}$ & $\tau_{14}$ & $\tau_{15}$ \\ 
				\midrule
				A1 & 1 & 3 & 5 & 2 & 4 & 1 & 1 & 2 & 5 & 5 & 10 & 8 & 2 & 4 & 2 \\ 
				A2 & 11 & 5 & 10 & 9 & 9 & 12 & 9 & 7 & 11 & 12 & 4 & 7 & 5 & 6 & 5 \\ 
				A3 & 6 & 10 & 8 & 11 & 11 & 8 & 11 & 8 & 12 & 9 & 6 & 11 & 12 & 11 & 11 \\ 
				A4 & 3 & 2 & 4 & 3 & 1 & 4 & 2 & 10 & 1 & 6 & 8 & 2 & 1 & 1 & 1 \\ 
				A5 & 9 & 4 & 7 & 5 & 6 & 6 & 6 & 5 & 3 & 3 & 2 & 5 & 11 & 7 & 9 \\ 
				A6 & 10 & 9 & 3 & 6 & 5 & 11 & 5 & 9 & 6 & 7 & 3 & 1 & 6 & 8 & 7  \\ 
				A7 & 8 & 8 & 11 & 7 & 12 & 9 & 12 & 11 & 8 & 10 & 7 & 9 & 8 & 12 & 12\\ 
				A8 & 4 & 1 & 1 & 4 & 3 & 2 & 4 & 4 & 2 & 1 & 1 & 6 & 3 & 2 & 6  \\ 
				A9 & 2 & 12 & 9 & 8 & 8 & 5 & 7 & 3 & 9 & 8 & 11 & 12 & 7 & 5 & 8 \\ 
				A10 & 7 & 11 & 6 & 10 & 10 & 7 & 8 & 6 & 7 & 11 & 9 & 3 & 10 & 9 & 4 \\ 
				A11 & 5 & 7 & 2 & 1 & 2 & 3 & 10 & 1 & 10 & 2 & 5 & 4 & 9 & 3 & 3 \\ 
				A12 & 12 & 6 & 12 & 12 & 7 & 10 & 3 & 12 & 4 & 4 & 12 & 10 & 4 & 10 & 10 \\ 
				\midrule
				H1 & 4 & 8 & 5 & 8 & 4 & 11 & 4 & 3 & 8 & 9 & 4 & 4 & 3 & 11 & 8 \\ 
				H2 & 1 & 2 & 4 & 5 & 2 & 7 & 2 & 2 & 1 & 2 & 1 & 1 & 2 & 2 & 1 \\ 
				H3 & 2 & 3 & 2 & 2 & 1 & 4 & 1 & 1 & 2 & 1 & 6 & 5 & 5 & 3 & 3 \\ 
				H4 & 3 & 4 & 3 & 4 & 3 & 3 & 3 & 4 & 3 & 4 & 7 & 7 & 1 & 1 & 2 \\ 
				H5 & 12 & 12 & 12 & 12 & 12 & 12 & 12 & 12 & 12 & 12 & 10 & 12 & 12 & 9 & 12 \\ 
				H6 & 6 & 5 & 1 & 1 & 6 & 2 & 7 & 5 & 7 & 3 & 5 & 3 & 7 & 4 & 6 \\ 
				H7 & 8 & 11 & 6 & 9 & 10 & 9 & 11 & 11 & 10 & 11 & 11 & 11 & 6 & 7 & 10 \\ 
				H8 & 11 & 6 & 8 & 3 & 7 & 1 & 6 & 6 & 6 & 6 & 8 & 8 & 4 & 8 & 9 \\ 
				H9 & 5 & 7 & 10 & 11 & 5 & 10 & 10 & 10 & 11 & 8 & 2 & 6 & 10 & 12 & 4 \\ 
				H10 & 10 & 9 & 9 & 7 & 9 & 5 & 5 & 7 & 5 & 7 & 12 & 9 & 11 & 5 & 7 \\ 
				H11 & 9 & 10 & 7 & 10 & 11 & 8 & 9 & 8 & 9 & 10 & 9 & 10 & 8 & 6 & 11 \\ 
				H12 & 7 & 1 & 11 & 6 & 8 & 6 & 8 & 9 & 4 & 5 & 3 & 2 & 9 & 10 & 5 \\ 
				\bottomrule
			\end{tabular}
	\end{table}
\begin{table}[ht]
	\scriptsize
	\centering
	\caption{Eleven covariates measured based on peer assessment rating (PAR) index. From left to right, the statistics stand for: Upper right segment; Upper anterior segment; Upper left segment; Lower right segment; Lower anterior segment; Lower left segment; Right buccal occlusion; Left buccal occlusion; Overjet; Overbit; Centerline.}
	\label{tab:ortho_covariate}
	\begin{tabular}{rrrrrrrrrrrr}
		\toprule
		& Urs & Uas & Uls & Lrs & Las & Lls & Rbo & Lbo & Oj & Ob & Cl \\
		\midrule
		A1 & 1.56 & 0.22 & 1.44 & 1.00 & 0.00 & 1.22 & 0.00 & 0.33 & 0.00 & 0.00 & 0.00 \\ 
		A2 & 1.33 & 0.22 & 1.00 & 0.33 & 0.00 & 0.33 & 0.00 & 0.33 & 0.00 & 0.33 & 0.00 \\ 
		A3 & 1.22 & 0.33 & 1.00 & 0.67 & 0.11 & 1.44 & 0.00 & 0.00 & 0.00 & 0.00 & 0.00 \\ 
		A4 & 0.00 & 0.00 & 0.11 & 1.78 & 0.22 & 1.89 & 0.33 & 0.67 & 0.00 & 0.00 & 0.00 \\ 
		A5 & 1.33 & 0.22 & 0.78 & 1.22 & 0.11 & 1.67 & 0.33 & 0.00 & 0.78 & 0.00 & 0.00 \\ 
		A6 & 1.11 & 0.56 & 1.78 & 0.89 & 0.22 & 0.89 & 0.67 & 1.00 & 0.78 & 0.00 & 0.00 \\ 
		A7 & 1.22 & 0.67 & 1.89 & 0.89 & 0.11 & 1.00 & 0.67 & 0.33 & 0.67 & 0.00 & 0.00 \\ 
		A8 & 1.44 & 0.22 & 1.56 & 0.89 & 0.22 & 0.56 & 2.00 & 2.00 & 0.00 & 0.00 & 0.00 \\ 
		A9 & 1.11 & 0.33 & 1.22 & 0.44 & 0.00 & 1.00 & 2.33 & 0.67 & 0.00 & 0.00 & 0.00 \\ 
		A10 & 0.67 & 0.11 & 0.89 & 0.11 & 0.00 & 0.00 & 0.67 & 1.00 & 0.00 & 0.67 & 0.00 \\ 
		A11 & 0.67 & 0.89 & 1.00 & 0.67 & 1.33 & 2.44 & 1.33 & 1.00 & 0.11 & 0.00 & 0.67 \\ 
		A12 & 0.67 & 0.11 & 0.22 & 1.00 & 0.00 & 0.56 & 0.33 & 1.33 & 0.00 & 0.33 & 0.00 \\ 
		\midrule
		H1 & 0.67 & 0.22 & 0.78 & 1.67 & 0.56 & 0.78 & 0.67 & 0.00 & 0.78 & 0.00 & 0.00 \\ 
		H2 & 1.56 & 0.56 & 0.22 & 0.44 & 0.00 & 0.11 & 0.00 & 0.67 & 0.00 & 0.00 & 0.00 \\ 
		H3 & 0.56 & 0.22 & 1.00 & 0.33 & 0.11 & 0.78 & 0.00 & 0.67 & 0.00 & 0.33 & 0.00 \\ 
		H4 & 0.56 & 0.22 & 0.67 & 0.44 & 0.11 & 0.44 & 0.67 & 1.00 & 0.00 & 0.00 & 0.00 \\ 
		H5 & 1.22 & 0.33 & 0.67 & 0.44 & 0.00 & 0.33 & 1.00 & 0.67 & 0.33 & 0.00 & 0.00 \\ 
		H6 & 0.56 & 0.11 & 1.33 & 1.22 & 0.00 & 1.33 & 1.00 & 0.67 & 0.22 & 0.00 & 0.00 \\ 
		H7 & 0.56 & 0.33 & 0.78 & 0.78 & 0.00 & 1.22 & 2.00 & 1.33 & 0.44 & 0.33 & 0.00 \\ 
		H8 & 0.78 & 0.22 & 1.56 & 0.89 & 0.00 & 0.33 & 1.67 & 2.00 & 0.00 & 0.00 & 0.00 \\ 
		H9 & 0.44 & 0.22 & 1.00 & 0.00 & 0.11 & 0.11 & 1.00 & 0.00 & 0.00 & 0.00 & 0.00 \\ 
		H10 & 1.11 & 0.33 & 1.78 & 0.22 & 0.22 & 0.33 & 1.33 & 1.67 & 0.00 & 0.00 & 0.00 \\ 
		H11 & 0.67 & 0.67 & 1.00 & 0.67 & 0.56 & 0.56 & 1.00 & 1.00 & 0.11 & 0.00 & 0.00 \\ 
		H12 & 1.22 & 0.78 & 1.00 & 0.33 & 0.33 & 0.67 & 1.00 & 0.67 & 0.56 & 0.00 & 0.00 \\ 
		\bottomrule
	\end{tabular}
\end{table}
\end{example}

Understanding how each orthodontist used the available covariates to arrive at his/her rank list and how to form a consensus ranking  are  important issues in this example, because the average perception of experienced orthodontists is considered the cornerstone of systems for the evaluation of orthodontic treatment outcome as described in \cite{song2014reliability}. However, Example \ref{eg:ortho} differs from Example \ref{eg:nfl} and prevailing rank aggregation applications in that it contains many ``local" rankings among non-overlapping subgroups. Having been demonstrated to be associated with ranking outcomes by \cite{song2015validation}, the covariate information can not only help generating a consensus full ranking list, but also  potentially reveal inhomogeneity among these experts in their ranking ``qualities'' as well as their way of using the covariates, 
\citep{liu2012consistency, song2014reliability}. As shown in Table \ref{tab:ortho_rank_data} and our later analysis, there are clearly heterogeneous qualities or opinions among rankers. For example, the ranking position of case A9 from the listed 15 experts in Table \ref{tab:ortho_rank_data} ranges from 2 to 12. 


\subsection{Key ideas and challenges}
There are mainly two types of methods dealing with rank data.
The first type  tries to find an aggregated ranking list that is consistent with most input rankings according to some criteria. For example, 
\citet{borda1781memoire} aggregated rankings based on the arithmetic mean of ranking positions, commonly known as Borda count.
  \citet{van2000variants} studied several variants of Borda count.
\citet{dwork2001rank} proposed to aggregate rankings based on 
the stationary distributions of certain Markov chains, which are constructed heuristically based on the ranking lists; and \citet{deconde2006combining} and \citet{lin2010rank} extended this approach to fit more complicated situations.
\citet{lin2009integration} obtained the aggregated ranking list by minimizing its total distance to all the input ranking lists, an idea that can be traced back to the Mallows model \citep{mallows1957non}. 
Recently, \citet{vitelli2017probabilistic} proposed an efficient MCMC approach to conduct Bayesian inference for the Mallows model and its extension allowing mixture heterogeneous subgroups of rankers, which naturally provides uncertainty quantification for the resulting quantities of interest. \citet{li2020mallows} formulated a different extension of the Mallows model and provided both  new theoretical results and an EM algorithm for the inference.
To overcome the computational difficulty and relax model assumptions, \citet{SVENDOVA2017122} proposed an indirect inference approach that tries to minimize the difference between the empirical distribution functions of entities' ranks and the corresponding true ones, and used non-parametric bootstrap to quantify uncertainty. 

The second type builds statistical models to characterize the data generating process of the rank data and uses the estimated models to generate the aggregated ranking list
\citep{blockmarschak1960, McFadden1980, Diaconis1988, critchlow1991probability, marden1996analyzing, alvo2014statistical}.  
The most popular model for rank data is  the Thurstone order statistics  model,
which includes the Thurstone--Mosteller--Daniels (TMD) 
model \citep{thurstone1927law, mosteller1951remarks, daniels1950rank} and Plackett--Luce (PL) model \citep{Bradley1952, luce1959, plackett1975} as special cases. 
Together with variants and extensions \citep{Stern1990, bockenholt1992thurstonian, WALKER2002303}, the Thurston model family has been successfully applied to a wide range of problems \citep[e.g.,][]{murphy2006, murphy2008jasa,  johnson2002bayesian, graydavies2016}.
Briefly, the Thurstone  model assumes that there is an underlying evaluation score for each entity, whose noisy versions determine the rankings. 
In the 
TMD 
and 
PL
models, the noises are assumed to follow the normal and Gumbel distributions, respectively. 
The  
PL 
model can be equivalently viewed as a multistage model that models the ranking process sequentially, where each entity has a unique parameter representing its probability of being selected at each stage up to a normalizing constant. 
A closely related literature to the development of rank data analysis is the analysis of pairwise comparison data; see \citet{Bradley1952} and \citet{david1963method} for earlier development, 
\citet{Davidson1976} and \citet{hastie1998} for applications, 
\citet{luce1959, Rao1967, plackett1975, Agresti1990} and \citet{Huang2006} for various extensions, 
and \citet{hunter2004mm, guiver2009bayesian, gormley2009} and \citet{Doucet2012} for efficient Bayesian computation.

Challenges arise in the analysis of ranking data when (a) rankers are of different qualities or belong to groups with different opinions; (b) covariate information are available for either the rankers or the ranked entities or both; and (c) there are incomplete ranking lists. 
\citet{murphy2006,murphy2008aoas,murphy2008jasa, murphy2010} developed the finite mixture of 
PL 
models and Benter models \citep{benter1994} to accommodate heterogeneous subgroups of rankers,  where both the mixing proportion and group-specific parameters can depend on the covariates of rankers. 
\citet{bockenholt1993} introduces the finite mixture of Thurstone models to 
allow for heterogeneous subgroups of rankers, with limited explorations;
\citet{yu2000bayesian} attempts to incorporate the covariate information for both ranked entities and rankers;  \citet{johnson2002bayesian} examines qualities of several known subgroups of rankers;  
and \citet{lee2014}  represents qualities of rankers by letting them  have different noise levels. See \citet{bockenholt2006} for a review of developments in Thurstonian-based analysis with some further extensions.  
In the presence of incomplete ranking lists, 
\citet{ailon2010aggregation, xia2011maximum, meila2012dirichlet} and \citet{liu2019learning} studied rank aggregation under various model assumptions. 
Recently, 
\citet{deng2014bayesian} proposed a Bayesian approach that can distinguish high-quality rankers from low-quality ones,  and 
\citet{bhowmik2017} proposed a method that utilizes covariates of ranked entities to assess qualities of all rankers. See also \citet{Badgeley2014,Wangaggregation2017,wangaggregation2018} for rank aggregation with application to genomic studies.


\subsection{The scope: a unified Bayesian framework and its extension}
Although various Thurstonian models have been proposed in the past, their inferences are mostly based on the maximum likelihood approach and the EM algorithm (with a few exceptions). The way of quantifying uncertainties and dealing with incomplete information has been limited.
In this paper,  we  propose a unified framework built upon the classic Thurstone  model family  to deal with incomplete ranking lists,  to accommodate rankers with different qualities or 
opinions, 
and to incorporate covariate information of ranked entities. 
In particular, 
we use the Dirichlet process prior for the mixture subgroups of rankers, which can automatically determine the total number of mixture components.  
Moreover, in addition to providing a full Bayesian inference procedure 
for  parameter estimation of the proposed models,   
we also pay special attention to rank aggregation and the uncertainty evaluation of the resulting aggregated ranking lists.

In this paper, we mainly focus on the TMD model and its extension to accommodate various complications. 
The Thurstone-type model is probably one of the most natural data generating model for rank data. 
It contains a rich family of statistical models including both the TMD model and the PL model, and is widely used in practice. 
We focus on the TMD model with  Gaussian errors since parametric Gaussian regression models are more intuitive and interpretable, are easier to manipulate in terms of model development, and have rich literature support.
Similar extensions can be made for other Thurstone models 
as well. 
The estimation for the 
TMD model is generally difficult due to the complicated form of the likelihood function, especially when there are a large number of ranked entities.  To overcome the difficulty, \citet{maydeu1999thurstonian} transformed the estimation problem to one involving mean and covariance structures with dichotomous indicators,  \citet{yao1999bayesian} proposed a Bayesian approach based on Gibbs sampler,  and 
\citet{johnson2013bayesian} advocated the JAGS software to implement the Bayesian posterior sampling. Our new model is even more challenging than the classic Thurstone family of models because of its inclusion of new components for dealing with   heterogeneous rankers. We  design an efficient parameter-expanded Gibbs sampler algorithm \citep{liu1999parameter}, which facilitates group moves of the latent variables and greatly improves the computational efficiency.  

Our extension of the 
TMD 
model is similar in spirit to \citet{murphy2006,murphy2008aoas,murphy2008jasa, murphy2010}'s extension of the 
PL 
model, but we allow infinite mixture components using the Dirichlet process prior. 
Moreover, unlike the 
PL
model, 
the 
TMD 
model does not have a closed-form likelihood, and thus impose additional computational challenges.
The rest of this article is organized as follows. Sections \ref{sec:latent_regression} and \ref{sec:extension} elaborate on our Bayesian models for rank data with covariates. Section \ref{sec3} provides details of our Markov Chain Monte Carlo (MCMC) algorithms. Section \ref{agg_mcmc} introduces multiple analysis tools using MCMC samples. Section \ref{sec:simulation} displays simulation results to validate our approaches. Section \ref{sec:realdata} describes the two real-data applications using the proposed methods. 
Section \ref{sec6} concludes with a short discussion.

\section{Latent regression models for rank data} 
\label{sec:latent_regression}
\subsection{Notation and definition}

Let $\mathcal{U}=\{1,2,\ldots, N\}$ be the set of all entities in consideration, and $N=|\mathcal{U}|$ be the total number of entities in $\mathcal{U}$. 
We use $i_1 \succ i_2$ to denote that entity $i_1$ is ranked higher than entity $i_2$. 
A \textit{ranking list} $\tau$ is  a set of non-contradictory pairwise relations in $\mathcal{U}$, which gives rise to an ordered preference list for 
entities in $\mathcal{U}$. 
We call $\tau$  a \textit{full ranking list} 
if $\tau$ identifies all pairwise relations in $\mathcal{U}$, otherwise a \textit{partial ranking list}. When $\tau$ is a full ranking list, we can equivalently write $\tau$ as $\tau=[i_1\succ i_2\succ \ldots \succ i_{N}]$ for notational simplicity, and further define $\tau(i)$ as the \textit{ranked position} of an entity $i\in \mathcal{U}$. 
Specifically, 
a higher ranked entity has a smaller numbered position in the list, i.e. $\tau(i_1)<\tau(i_2)$ if and only if $i_1 \succ i_2$. 
For example, Tables \ref{tab:NFL_rank_data} and \ref{tab:ortho_rank_data} show the ranked positions of the entities in each ranking list. 
Furthermore, for any vector $\bm{z} = (z_1, \ldots, z_N)^\top \in \mathbb{R}^N$, we use $\text{rank}(\bm{z}) = [i_1\succ i_2\succ \ldots \succ i_{N}]$ to denote the full ranking list of $z_i$'s in a decreasing order, i.e., $z_{i_1} \geq \ldots \geq z_{i_N}$. 

As introduced in Examples \ref{eg:nfl} and \ref{eg:ortho}, we also observe some covariates of ranked entities. 
Let $\bm{x}_i \in \mathbb{R}^L$ be the $L$ dimensional covariate vector of ranked entity $i$, and 
$\bm{X}=(\bm{x}_1,\bm{x}_2,\ldots,\bm{x}_N)^\top \in \mathbb{R}^{N\times L}$ be the covariate matrix for all $N$ entities. 
In the remaining discussion, 
for clarification, we use index $1\leq i\leq N$ for ranked entities and index $1 \leq j \leq M$ for rankers, with $N$ and $M$  denoting the total numbers of ranked entities and rankers, respectively.

\subsection{The Thurstone--Mosteller--Daniels model 
	without covariates}\label{sec:TMD_no_covariate}

Suppose we have $M$ full ranking lists $\tau_1,\tau_2,\ldots,\tau_M$ for entities in $\mathcal{U}=\{1, 2,$ $\ldots,N\}$. 
\citet{thurstone1927law} postulated that the ranking outcome $\tau_j$ is determined by $N$ latent variables $Z_{ij}$'s for $1\leq i \leq N$, where $Z_{ij}$ represents ranker $j$'s evaluation score of the $i$th entity, 
and $Z_{i_1j}>Z_{i_2j}$ if and only if $i_1 \succ i_2$ for ranker $j$.
Define $\bm{Z}_j=(Z_{1j},\ldots,Z_{Nj})^\top$ as ranker $j$'s evaluations of all entities, 
and 
$\text{rank}(\bm{Z}_j)$ as the associated full ranking list based on $\bm{Z}_j$. 
Under the TMD model, 
$\bm{Z}_j$ follows a multivariate Gaussian distribution 
with mean $\bm{\mu}=(\mu_1,\ldots,\mu_N)'$ representing the underlying true score of the ranked entities:
\begin{align}\label{eq:BAR}	
Z_{ij} & =\mu_{i}+\varepsilon_{ij}, \ \quad \varepsilon_{ij}\sim N(0,\sigma^2) & (1\leq i\leq N; 1\leq j\leq M)\\
\tau_j & =\text{rank}(\bm{Z}_j), &  (1\leq j\leq M)
\nonumber
\end{align}
where $\varepsilon_{ij}$'s are mutually independently across all $i$ and $j$. 
Because we only observe the ranking lists $\tau_j$, 
multiplying $(\bm{\mu},\sigma)$ by a constant or adding a constant to all the $\mu_i$'s does not influence the likelihood function.
Therefore, to ensure identifiability of the parameters, 
we fix $\sigma^2=1$ and impose the constraint that  $\bm{\mu}$ lies in the space $\bm{\Theta}=\{\bm{\mu}\in \mathbb{R}^n:\bm{1}_N^\top \bm{\mu}=0\}$, 
where $\bm{1}_N$ is an $N$ dimensional vector with all coordinates being 1. 

Model \eqref{eq:BAR} implies that the $\tau_j$'s are independent and identically distributed (i.i.d.) conditional on $\bm{\mu}$. The likelihood function is then $p(\tau_1, \tau_2, \ldots,\tau_M \mid \bm{\mu})=\prod_{j=1}^M p(\tau_j \mid \bm{\mu})$ with 
\begin{align}\label{eq:prob_tau}
	p(\tau_j \mid \bm{\mu})
	& =  \int_{\mathbb{R}^N} p(\tau_j \mid \bm{Z}_j,\bm{\mu}) p(\bm{Z}_j \mid \bm{\mu})\text{d}\bm{Z}_j\\
	& = \int_{\mathbb{R}^N} 1_{\{\text{rank}(\bm{Z}_j)=\tau_j\}}
	\cdot (2\pi)^{-N/2}e^{-\|\bm{Z}_j-\bm{\mu}\|^2/2} 
	\text{d}\bm{Z}_j.
	\nonumber
\end{align}
The goals are to 
estimate the parameter $\bm{\mu}$ and then 
generate an aggregated ranking based on the estimated $\bm{\mu}$. 
A common approach is the maximum likelihood method, which is computationally challenging due to the integral in the likelihood. 
Besides, it is also nontrivial to quantify the uncertainty of the resulting rank aggregation. 
We focus on the Bayesian approach,
%
which is more convenient to incorporate  prior information, to quantify estimation uncertainties, and  to utilize efficient 
MCMC algorithms including 
data augmentation \citep{tanner1987calculation} and parameter expansion strategies \citep{liu1999parameter}. 
With a reasonable prior on the $\mu_i$'s, 
we can get the corresponding posterior means 
of $\mu_i$'s, based on which we can generate an aggregated ranking list. 

Recalling that $\bm{\mu}$ is restricted to the space $\bm{\Theta}$, 
we define $\bm{P}_N = \bm{I}_N - N^{-1}\bm{1}_N\bm{1}_N^\top$ as the projection matrix that maps any vector in $\mathbb{R}^N$ to $\bm{\Theta}$, 
where $\bm{I}_N$ is an $N \times N$ identity matrix and $\bm{1}_N$ is an $N$ dimensional vector with all elements being 1. 
We choose the prior of $\bm{\mu}$
to be $\mathcal{N}\left(\bm{0},\sigma_{\mu}^2\bm{P}_N\right)$. The intuition for choosing this prior is that when $\bm{\mu}\sim \mathcal{N}(\bm{0}, \sigma_{\mu}^2 \bm{I}_N)$, we have $\bm{P}_N\bm{\mu}\in \bm{\Theta}$ and $\bm{P}_N \bm{\mu} \sim \mathcal{N}(\bm{0},\sigma_{\mu}^2\bm{P}_N)$. 
For computational convenience, it is equivalent to using the prior $\bm{\mu}\sim \mathcal{N}(\bm{0}, \sigma_{\mu}^2 \bm{I}_N)$ and considering the posterior mean of $\bm{P}_N\bm{\mu} \equiv \bm{\mu}-\bar{\mu}\bm{1}_N$, where $\bar{\mu}=n^{-1}\sum_{i=1}^{n}\mu_i$. 
In other words,
$
f_{\pi_1}(\bm{\mu}\mid\tau_1,\cdots,\tau_M) = f_{\pi_2}( \bm{\mu}-\bar{\mu}\bm{1}_N \mid\tau_1,\cdots,\tau_M),
$
where $\pi_1(\bm{\mu})\sim \mathcal{N}\left(\bm{0},\sigma_{\mu}^2\bm{P}_N\right)$ and $\pi_2 (\bm{\mu}) \sim \mathcal{N}(\bm{0}, \sigma_{\mu}^2 \bm{I}_N)$ denote the priors of $\bm{\mu}$. More generally, although we restrict $\bm{\mu}$ to the parameter space $\bm{\Theta}$, we only need to specify a prior for the unconstrained $\bm{\mu}$ and 
make inference based on the posterior distribution of $\bm{\mu}-\bar{\mu}\bm{1}_N$.
Therefore, 
in the following discussion, we relax the constraint that $\bm{\mu}\in \bm{\Theta}$, 
and emphasize that what matters are the relative values of the $\mu_i$'s. 

\subsection{The Thurstone--Mosteller--Daniels model with covariates}\label{sec:tmd_cov}
As in both examples, each ranked entity is associated with some covariates that may be 
relevant to how the entity is perceived and thus ranked by rankers. 
To incorporate the covariate information into model \eqref{eq:BAR}, we assume that the score of each entity $i$  depends linearly on the $L$-dimensional covariate vector $\bm{x}_i$, for $i=1,\ldots, N$.  To avoid being too restrictive, we allow the intercept term for each entity to be different.
In particular, we have the following over-parameterized model: 
\begin{align}\label{eq:BARC}	
\mu_{i} &=\alpha_{i}+\bm{x}_{i}^\top \bm{\beta},
& (1\leq i\leq n)
\nonumber
\\
Z_{ij} &=\mu_{i}+\varepsilon_{ij}, \quad  \varepsilon_{ij}\sim \mathcal{N}(0,1), \quad 
& (1\leq i\leq N; 1\leq j\leq M)\\
\tau_j &=\text{rank}(\bm{Z}_j), & (1\leq j\leq M) \nonumber
\end{align}
where the $\varepsilon_{ij}$'s are jointly independent across all $i$ and $j$.

Model \eqref{eq:BARC} is over-parameterized because $\bm{\mu}$ 
is invariant if we add a constant vector $\bm{c}$ to $\bm{\beta}$ and change $\alpha_i$ to $\alpha_i-\bm{x}_{i}^\top \bm{c}$. 
However, 
the structure between $\bm{\mu}$ and $(\bm{\alpha},\bm{\beta})$ can help us construct some informative priors on $\bm{\mu}$, incorporating the covariate information. 
Intuitively, entities with similar  $\bm{x}_{i}$'s should be close in the underlying $\mu_i$'s. 
Such intuition is conformed by model \eqref{eq:BARC} with suitable priors on $(\bm{\alpha},\bm{\beta})$, 
because similar entities will have higher correlation among their $\mu_i$'s \textit{a priori}.   
Model \eqref{eq:BARC} can be helpful when the ranking information is weak and incomplete, and the covariate information is strongly related to the ranking mechanism. More generally, some covariates of the rankers may also be available, and they can be similarly incorporated into the $\mu_i$'s in \eqref{eq:BARC}.
For example, with covariates $\bm{v}_j$ for each ranker $j$, we can model the evaluation score of ranker $j$ for entity $i$ as $Z_{ij} = \alpha_i + \bm{x}_i^\top \bm{\beta} + \bm{v}_j^\top \bm{\delta}_i+\varepsilon_{ij}$, similar to \citet{yu2000bayesian} and \citet{murphy2010}. 
In this paper, we focus only on 
the covariates of the ranked entities mainly due to our applications. 

We further illustrate model \eqref{eq:BARC} using the quarterback data in Example \ref{eg:nfl}. 
The unobserved variable $Z_{ij}$ represents ranker $j$'s evaluation for the performance of quarterback $i$. The expression $\alpha_{i}+\bm{x}_{i}^\top \bm{\beta}$ quantifies a hypothetically universal underlying ``score'' of the quarterback, and each ranker evaluates it with a personal variation modeled by $\varepsilon_{ij}$. The linear term $\bm{x}_{i}^\top \bm{\beta}$ can explain part of their performance, but there are many aspects in a football game that cannot be reflected through a linear combination of these summary statistics. The term $\alpha_i$ can capture the remaining ``random effect''. Without $\alpha_i$, model \eqref{eq:BARC} reduces to a rank regression model in \citet{johnson2013bayesian}, which can be too restrictive in some applications.

We set the prior $\pi(\bm{\alpha},\bm{\beta})\equiv \pi(\bm{\alpha}) \pi(\bm{\beta})$, where $\pi(\bm{\alpha})$
is $\mathcal{N}(0,\sigma^2_\alpha \bm{I}_{N})$ and $p(\bm{\beta})$ is $\mathcal{N}(0,\sigma^2_\beta \bm{I}_L)$, where $\bm{I}_{N}$ and $\bm{I}_L$ are $N\times N$ and $L\times L$ identity matrices, respectively. 
The hyper-parameter $\sigma_\alpha$ and $\sigma_\beta$ can reflect our prior belief on the relevance of covariate information to ranking mechanism. 
Intuitively, the stronger the belief on the role of covariates, the smaller the ratio $\sigma^2_{\alpha}/\sigma^2_{\beta}$ should be chosen. 
The Bayesian procedure based on the generalized regression model \eqref{eq:BARC} is henceforth referred to as Bayesian Analysis of Rank data with Covariates (BARC). 

\subsection{The Plackett-Luce model}\label{sec:PL_model}
The PL model is another popular Thurstone model for rank data, and it 
differs from the TMD model described in Sections \ref{sec:TMD_no_covariate} and \ref{sec:tmd_cov} in the distributional assumption for the noises $\varepsilon_{ij}$'s. 
In particular, the PL model assumes that all the noises $\varepsilon_{ij}$'s follow i.i.d. Gumbel distribution. 
Importantly, the probability of observing a rank list $\tau_j$ under the parameter $\bm{\mu}$, which can be written as an integral as in \eqref{eq:prob_tau}, now has an equivalent closed-form expression: 
\begin{align}\label{eq:prob_tau_PL}
	p(\tau_j \mid \bm{\mu}) & = \prod_{i=1}^N \frac{\exp\left( \mu_{\tau_j(i)} \right) }{\sum_{i' = i}^N \exp\left( \mu_{\tau_j(i')} \right) }. 
\end{align}
Therefore, 
compared to the TMD model, 
the PL model requires less computational cost, because the likelihood for the observed rankings has a closed-form expression that does not involve any integral. 
\citet{hunter2004mm} proposed a minorization-maximization (MM) algorithm for finding the MLE of the parameters,  \citet{guiver2009bayesian} worked out a Bayesian approach based on a message-passing algorithm (Expectation-Propagation),
\citet{Doucet2012} proposed simple Gibbs samplers for Bayesian inference, 
and \citet{NIPS2013_4997} proposed a class of  generalized method-of-moments algorithms. 
The covariate information can be similarly included as in \eqref{eq:BARC}, 
see, e.g., \citet{HAUSMAN198783,Allison1994}. 

The performance of the TMD model and the PL model depend on the nature of the data, and thus is generally case-by-case; see \citet[][Chapter 2]{azari2014} for a simulation study. 
In this paper, we mainly focus on the TMD model, and the implementation can be helpful for general noise distributions that may not be easily integrated out.

\section{Extension of the latent regression model}\label{sec:extension}
\subsection{Rankers with varying qualities}
In practice, the rankers in consideration 
may have different quality or reliability. 
In these cases, 
it is of interest to distinguish high-quality rankers from low-quality ones, 
and
a weighted rank aggregation method is often preferred, 
where each ranker $j$ has a weight $w_j$ 
reflecting the quality of its ranking list.
However, it is generally difficult to  design a proper weighting scheme in practice, especially when little or no prior knowledge of the rankers is available. 
To deal with this difficulty, 
one can accommodate weighting through the variance parameters in the model, and infer them jointly with other parameters. More precisely, we model the ranker's quality  by  the precision of the noise, i.e, extending  model \eqref{eq:BARC} to the following weighted version:
\begin{align}\label{eq:BARCW}
\mu_i & = \alpha_{i}+\bm{x}_{i}'\bm{\beta}, & (1\leq i\leq N)
\nonumber
\\
Z_{ij} & =\mu_i+\varepsilon_{ij}, \ 
\quad \varepsilon_{ij}\sim N(0,w^{-1}_j), & (1\leq i\leq N; 1\leq j\leq M)\\
\tau_j & =\text{rank}(\bm{Z}_j), & (1\leq j\leq M)
\nonumber
\end{align}
where $w_j>0$ for all $j$ and the $\varepsilon_{ij}$'s are mutually independent across all $i$ and $j$.

The prior for the  $w_j$'s can be any
distribution with support on positive real numbers, such as uniform and truncated chi-squared distributions. 
In the absence of covariates, \citet{lee2014} considered model \eqref{eq:BARCW} and assumed that the $w_j^{1/2}$'s are i.i.d. uniform on $(0,20)$ {\it a priori}. 
Here we consider a more restrictive choice where the weights take on only a few values, 
which can lead to a 
less sticky MCMC sampler without compromising much  precision in the rankers' quality evaluation and 
the aggregated ranking list. 
Specifically, we restrict $w_j$ to three values, 2, 1 and 0.5, standing for reliable, mediocre, and low-quality rankers, respectively, with equal probabilities {\it a priori}, i.e.,
\begin{equation}
P(w_j=0.5)=P(w_j=1)=P(w_j=2)= 1/3, \quad (1\leq j\leq M)
\end{equation}
and assume 
the $w_j$'s are mutually independent across all $j$. 
We call the resulting rank analysis method  the Bayesian Analysis of Rank data with entities' Covariates and rankers' (unknown) Weights (BARCW, henceforth).

\subsection{Ranker clustering via mixture model}
All previously described models assume that the underlying score $\bm{\mu}$ is universal to all rankers, which can sometimes be too restrictive. \cite{bockenholt1993} and \cite{murphy2006,murphy2008aoas,murphy2008jasa}  suggested that there are often several categories of voters with very different political opinions in an election, and subsequently a mixture model approach should be applied to cluster voters into subgroups. Differing from BARCW, which studies differences in rankers' reliability, this mixture model focuses on the heterogeneity in rankers' opinions while assuming that all rankers are equally reliable.

\cite{bockenholt1993} considered finite mixtures of TMD models.
However, in finite mixture models, 
a common issue is how to determine the number of mixture components. 
Here we employ the Dirichlet process mixture model, which
overcomes this issue by using mixture distributions with countably infinite number of components via a Dirichlet process prior \citep{antoniak1974mixtures, ferguson1983bayesian}. 
We first extend model \eqref{eq:BARC} so that the underlying score of each entity is ranker-specific:
\begin{equation}
\begin{aligned}
\bm{\mu}^{(j)} &= \bm{\alpha}^{(j)}+\bm{X}\bm{\beta}^{(j)}, & \quad (1\leq j\leq M)\\
\bm{Z}_{j} & = \bm{\mu}^{(j)} + \bm{\varepsilon}_j, \  \quad 
\bm{\varepsilon}_j \sim  \mathcal{N}(\bm{0},\bm{I}_n), & \quad  (1\leq j\leq M)\\
\tau_{j} &= \text{rank}\left(\bm{Z}_{j}\right), & \quad (1\leq j\leq M)
\end{aligned}
\label{eq:BARCM1}
\end{equation}
where $\bm{X} \in \mathbb{R}^{N\times L}$ is the covariate matrix for all ranked entities, 
$\bm{\mu}^{(j)}$ represents the underlying true score 
for ranker $j$, and $\bm{\varepsilon}_j$'s are mutually independent. 
We then assume that the  $(\bm{\alpha}^{(j)},\bm{\beta}^{(j)})$ follow a distribution $G$ that is drawn from a Dirichlet process, i.e.,
\begin{align}\label{eq:BARCM2}
(\bm{\alpha}^{(j)},\bm{\beta}^{(j)})\mid G \overset{iid}{\sim} G, \quad 
G &\sim DP(\gamma,G_0),
\end{align}
where $G_0$ defines a baseline distribution on $\mathbb{R}^{N+L}$, 
and $\gamma$ is a concentration parameter. 
For the ease of understanding, we can equivalently view  model \eqref{eq:BARCM1}-\eqref{eq:BARCM2} as the limit of 
the following finite mixture model with $K$ components when $K\rightarrow \infty$:
\begin{align*}
( \bm{\alpha}^{\langle k\rangle},\bm{\beta}^{\langle k\rangle} ) &\overset{i.i.d.}{\sim} G_0, &   (1 \leq k \leq K)\\
\bm{\mu}^{\langle k\rangle} & = \bm{\alpha}^{\langle k\rangle}+\bm{X}\bm{\beta}^{\langle k\rangle}, 
&   (1 \leq k \leq K)
\nonumber\\
\left(\pi_1, \ldots, \pi_K\right) & \sim \text{Dir}(\gamma/K, \ldots, \gamma/K), 
\nonumber\\
c_{j} \mid \bm{\pi} &\overset{i.i.d.}\sim \text{Multinomial}\left(\pi_1, \ldots, \pi_K\right), & (1\leq j\leq M)
\nonumber\\
\bm{Z}_j & = \bm{\mu}^{\langle c_j\rangle} + \bm{\varepsilon}_j, \quad \bm{\varepsilon}_j \sim \mathcal{N}(\bm{0}, \bm{I}_n), & \quad (1\leq j\leq M)
\nonumber\\
\tau_{j} & = \text{rank}\left(\bm{Z}_{j}\right), & \quad (1\leq j\leq M)
\nonumber
\end{align*}
where the latent variable $c_j\in \{1,2,\ldots, K\}$ indicates the cluster allocation of ranker $j$, and $\bm{\mu}^{\langle k\rangle}$ corresponds to the common underlying score vector for rankers in cluster $k$. 

We choose the baseline distribution $G_0$ on $\mathbb{R}^{N+L}$ using two independent zero-mean Gaussian distributions with covariances $\sigma_{\alpha}^2\bm{I}_N$ and $\sigma_{\beta}^2\bm{I}_L$, i.e., 
$G_0 \sim \mathcal{N}(\bm{0},$ $\text{diag}(\sigma_{\alpha}^2\bm{I}_N, \sigma_{\beta}^2\bm{I}_L))$. 
Section~\ref{sec:hyperpara} provides more details on how the prior for these parameters in the model might be set.
Obviously, 
$G_0$ 
is the same as the prior distribution of $(\bm{\alpha},\bm{\beta})$ we use in the previous models, and the conjugacy between $G_0$ and the distribution of $\bm{Z}_j$'s leads to a straightforward Gibbs sampler  \citep{neal1992bayesian,liu1994collapsed, maceachern1994estimating}. Parameter $\gamma$ represents the degree of concentration of $G$ around $G_0$ and is  related to the number of distinct clusters. According to the P{\'o}lya urn scheme representation of the Dirichlet process \citep{blackwell1973ferguson}, 
the expected number of clusters with in total $M$ rankers is $\sum_{j=1}^M\gamma/(j+\gamma-1)$ {\it a priori}. We discuss the sensitivity of these hyper-parameters in the simulation studies.

Under this Dirichlet process mixture model, we are interested in understanding the heterogeneous opinions among rankers and  
rank aggregation within each cluster as well as 
across all clusters. 
The aggregated ranking in each cluster 
is determined by those $\bm{\mu}^{(j)}$'s with identical values. 
The aggregated ranking list across all clusters depends on the underlying score of all rankers, i.e., 
$M^{-1}\sum_{j=1}^M \bm{\mu}^{(j)}$. 
We regard this rank analysis 
method as BARCM, standing for Bayesian Analysis of Rank data with Covariates of entities and Mixture of rankers with different 
opinions. 
Furthermore, it is also straightforward to further incorporate varying weights for all rankers as in BARCW using model \eqref{eq:BARCW}. 
To avoid being too lengthy, we skip the detailed description for the mixture model with varying weights, and simply denote it as BARCMW, standing for Bayesian Analysis of Rank data with Covariates of entities and Mixture of rankers with different opinions and Weights. 

\subsection{Extension to partial ranking lists}\label{sec:partial_list}

Models \eqref{eq:BAR}, \eqref{eq:BARC}, \eqref{eq:BARCW} and \eqref{eq:BARCM1}-\eqref{eq:BARCM2} can all be 
extended to cases where 
the observations are partial ranking lists. Because we define a ranking list as a set of non-contradictory pairwise relations among ranked entities, partial ranking lists appear when any of the pairwise relations is missing. Thus, besides the partial ranking list $\tau_j \ (1\leq j\leq M)$, we also observe the $\delta_j$'s that indicate which pairwise relationship is missing. 
Under the latent variable models, we denote  $\tau_j \simeq \text{rank}(\bm{Z}_j)$ if the partial ranking list $\tau_j$ is consistent with the full ranking list $\text{rank}(\bm{Z}_j)$. 
Our models, BARC, BARCW and BARCM, for the observed individual partial ranking lists
are the same as in \eqref{eq:BAR}, \eqref{eq:BARC}, \eqref{eq:BARCW} and \eqref{eq:BARCM1}-\eqref{eq:BARCM2}, except that $\tau_j = \text{rank}(\bm{Z}_j)$ is replaced by $\tau_j \simeq \text{rank}(\bm{Z}_j)$. 
Let $\bm{\theta}_\delta$ and $\bm{\theta}_\tau$ denote the parameters for missing indicators $\delta_j$'s and ranking lists $\tau_j$'s, respectively. 
We can then write the likelihood of $(\delta_j,\tau_j)$ as
\begin{align*}
p(\delta_j, \tau_j \mid \bm{\theta}_\delta, \bm{\theta}_\tau, \bm{X}) 
= 
\sum_{r: r \simeq \tau_j} 
\int_{\mathbb{R}^N}
p(\delta_j \mid r, \bm{Z}_j, \bm{\theta}_\delta, \bm{X}) 
1_{\{r = \text{rank}(\bm{Z}_j) \}}
p(\bm{Z}_j \mid \bm{\theta}_\tau, \bm{X})
\text{d} \bm{Z}_j.
\end{align*}
If the pairwise relations are missing at random, in the sense that
$p(\delta_j \mid r, \bm{Z}_j, \bm{\theta}_\delta, \bm{X})$ $= p(\delta_j \mid \tilde{r}, \tilde{\bm{Z}}_j, \bm{\theta}_\delta, \bm{X})$
for all possible $(r, \bm{Z}_j,  \tilde{r}, \tilde{\bm{Z}}_j)$ such that $r = \text{rank}(\bm{Z}_j) \simeq \tau_j$
and 
$\tilde{r}= \text{rank}(\tilde{\bm{Z}}_j) \simeq \tau_j$, then the likelihood of  $(\delta_j,\tau_j)$ can be simplified as 
\begin{align*}
p(\delta_j, \tau_j \mid \bm{\theta}_\delta, \bm{\theta}_\tau, \bm{X}) 
= p(\delta_j \mid \tau_j, \bm{\theta}_\delta, \bm{X}) 
\int_{\mathbb{R}^N}
1_{\{\tau_j \simeq \text{rank}(\bm{Z}_j) \}}
p(\bm{Z}_j \mid \bm{\theta}_\tau, \bm{X})
\text{d} \bm{Z}_j
\end{align*}
If further the priors for the parameters $\bm{\theta}_\delta$ and $\bm{\theta}_\tau$ are mutually independent, we can ignore the $\delta_j$'s when conducting Bayesian inference for the parameter $\bm{\theta}_\tau$ of ranking mechanisms. 


Here we give two additional remarks. 
First, we consider 
a special type of partial list, the top-$K$ list, from which we can observed only the top $K$ entities in a ranking list; see 
e.g., \citet{topklists2015} for more detailed discussion. 
When $K$ is fixed, it is not difficult to see that the corresponding pairwise comparison induced from a top-$K$ list is missing at random. 
Second, we consider rank data containing ties. 
Generally, under the Thurstone-type model with continuous errors, the ranking list will have ties with zero probability. 
Practically, to mitigate this issue, we may view observed ranking lists with ties as partial ranking lists. More explicitly, we may treat ties as missing pairwise comparisons. 
However, such missing is not at random, implying that treating the information contained in ``regarding the two entities as a tie" the same as ``providing no comparisons between the two  entities" may incur a little information loss, though it may not be of any practical importance.

\subsection{Extension of the Plackett-Luce model}

The discussion above mainly focuses on the extension of the TMD model with normal noise. 
Similar extension can also be made for the PL model with Gumbel noise. 
\citet{murphy2006} extended the PL model to allow a finite mixture of PL models. 
They further extended the model to mixture of experts, allowing the dependence of mixture probabilities on the ranker's covariates \citep{murphy2008aoas}, as well as the dependence of ranking mechanism on the ranker's covariates \citep{murphy2010}.  
A unique feature of the PL model is that it can be viewed as a multistage model as in \eqref{eq:prob_tau_PL}.  
\citet{benter1994} extended this multistage model by allowing the probability ratios of being top  among remaining entities to vary across different stages. 
Intuitively, 
under Benter's model,  
the choice of the top entity becomes increasingly random along the stages, and an entity with a smaller $\mu_i$ in \eqref{eq:prob_tau_PL} has a greater probability to be ranked at a higher position. 
\citet{murphy2008aoas,murphy2008jasa} also studied the extension of Benter's model to mixture model and mixture of experts. 
For these finite mixture models, the number of mixture components are selected based on some information criteria such as BIC. 
Furthermore, \citet{murphy2010} studied the choice of these models, including mixture model and mixture of experts, also using some model selection criteria, and illustrates how the ranker's covariate information should be incorporated into the PL model in practice. 
Recently, \citet{liu2019learning} studied the PL mixture model for partial ranking lists, and proposed MCMC-based computation tools.  

Our discussion for the TMD model involves only the covariates of the ranked entities, mostly due to our application. 
The covariate information of the rankers, if available, can also be incorporated similarly as in \citet{murphy2010}. 
Again, we focus mainly on the computation for the TMD model, which can be useful for general noise distributions that are difficult to integrate out.

\section{MCMC computation with parameter expansion}\label{sec3}

We advocate the use of Gibbs sampler with parameter expansion \citep{liu1999parameter} for Bayesian inference with general latent variable models, and in particular  the class of TMD-based models we introduced in the previous sections. 
The parameter-expansion idea has been applied to  
ordered data and rank data analysis in previous studies
\citep[e.g.,][]{liu1999parameter,hoff2009first,Fong2016}. 
Here we provide a unified parameter-expanded Gibbs sampling algorithm for the TMD model with rankers of varying qualities or heterogeneous opinions. 
We start with model \eqref{eq:BARC}
and then generalize this MCMC strategy to the extended models \eqref{eq:BARCW} and \eqref{eq:BARCM1}-\eqref{eq:BARCM2}. 
We also provide an R package for implementing the proposed Bayesian analysis, with detailed information relegated to the Supplementary Material. 
For notational convenience, 
we define $\bm{Z} = (\bm{Z}_1,\ldots,\bm{Z}_M)\in \mathbb{R}^{N\times M}$,  $\Tau = \{\tau_j\}_{j=1}^M$, 
$\bm{V} = (\bm{I}_N,\bm{X}) \in \mathbb{R}^{N\times (N+L)}$, and 
$\bm{\Lambda} = \text{diag}(\sigma_{\alpha}^{2}\bm{I}_{N}, \sigma_{\beta}^{2}\bm{I}_{L}) \in \mathbb{R}^{(N+L)\times(N+L)}$. 

\subsection{Parameter-expanded Gibbs sampler}
\label{step_zba}

The most computationally expensive part in our model is to sample all the $Z_{ij}$'s from the truncated Gaussian distributions. 
Moreover, because $\bm{Z}$ and $(\bm{\alpha}, \bm{\beta})$ are intertwined together due to the posited regression model, they tend to correlate highly, similar to the difficulty in the  data augmentation method 
for probit regression models \citep{albert1993bayesian}.

To speed up the convergence of the MCMC algorithm, we follow Scheme 2 in 
\citet{liu1999parameter} and exploit a parameter-expanded data augmentation (PX-DA) algorithm. 
In particular, we introduce a group scale transformation of the ``missing data'' matrix $\bm{Z}$, which contains the evaluation scores of all rankers for all entities, indexed by a 
positive 
parameter $\theta$, i.e., $t_{\theta}(\bm{Z})\equiv \bm{Z}/\theta$. 
For $1\le i \le N$ and $1\le j \le M$, 
let 
$\bm{Z}_{-j}$ denote the evaluation scores of all rankers except ranker $j$, 
and 
$\bm{Z}_{-i, j}$ denote ranker $j$'s evaluation scores of all entities except entity $i$. 
The PX-DA algorithm updates the missing data $\bm{Z}$ and the expanded parameters $(\theta, \bm{\alpha},\bm{\beta})$ iteratively as follows: 
\begin{enumerate}[label=(\arabic*)]
	\item[(i)] For $j=1,\ldots,M$ and $i=1,\ldots,N$, draw [$Z_{ij} \mid \bm{Z}_{-i,j},\bm{Z}_{-j},\bm{\alpha},\bm{\beta}$] from truncated  $\mathcal{N}(\alpha_i+\bm{x}_{i}'\bm{\beta},1)$, where the truncation points are determined by $\bm{Z}_{-i,j}$ and $\tau_j$
	such that $\text{rank}(\bm{Z}_j)\simeq\tau_j$. 
	
	\item[(ii)] Draw $\theta\sim p(\theta \mid \bm{Z},\Tau)\propto p(t_\theta(\bm{Z}))|J_\theta(\bm{Z})|H(d\theta)$, 
	and then update 
	$\bm{Z}$ to be $t_{\theta}(\bm{Z})$. 
	Here, $|J_\theta(\bm{Z})|=\theta^{-nm}$ is the Jacobian of scale transformation, 
	$H(d\theta)=\theta^{-1}d\theta$ is the Haar measure on a scale group up to a constant, 
	and
	$$
	p(t_\theta(\bm{Z})) \propto \int p(t_\theta(\bm{Z}) \mid \bm{\alpha}, \bm{\beta})p(\bm{\alpha})p(\bm{\beta})\text{d}\bm{\alpha} \text{d}\bm{\beta} \propto \exp\left\{-\frac{S}{2\theta^2}\right\},
	$$  
	is the marginal density of latent variables evaluated at $t_{\theta}(\bm{Z})$, 
	where 
	$$
	S=\sum_{j=1}^{M}\bm{Z}_j^\top \bm{Z}_j-\sum_{j=1}^{M}\sum_{j'=1}^{M}\bm{Z}_j^\top \bm{V}(\bm{\Lambda}^{-1}+M\bm{V}^\top \bm{V})^{-1}\bm{V}^\top\bm{Z}_{j'}.
	$$
	We can derive that $\theta^2\sim S/\chi^2_{NM}$. 
	\item[(iii)]	
	Draw $(\bm{\alpha},\bm{\beta})\sim p(\bm{\alpha}, \bm{\beta} \mid \bm{Z}) \sim  \mathcal{N}( \bm{\eta}, \bm{\Sigma} )$, 
	where
	$$
	\bm{\eta}=(\bm{\Lambda}^{-1}+M\bm{V}^\top \bm{V})^{-1}\bm{V}^\top \sum_{j=1}^M \bm{Z}_j
	\ \ 
	\text{  and  } 
	\ \ 
	\bm{\Sigma} = (\bm{\Lambda}^{-1}+M\bm{V}^\top \bm{V})^{-1}.
	$$
\end{enumerate}

Below we give some intuition on why the PX-DA algorithm improves efficiency. 
Without Step (ii), 
the algorithm reduces to the standard Gibbs sampler, which  updates the missing data and parameters iteratively. The scale group move of $\bm{Z}$ under the usual Gibbs sampler is slow due to both the Gibbs update for $\bm{Z}$ in Step (i) and the high correlation between $\bm{Z}$ and $(\bm{\alpha}, \bm{\beta})$. 
To overcome such difficulty, 
the PX-DA algorithm introduces a scale transformation of $\bm{Z}$ to facilitate its group move based on its marginal conditional distribution with $(\bm{\alpha}, \bm{\beta})$ integrated out. Thus, together with Step (iii), PX-DA effectively achieves the conditional sampling of
$(\bm{\alpha}, \bm{\beta})$ and a scale group move of $\bm{Z}$ jointly. 
To ensure the validity of the MCMC algorithm, the scale transformation parameter $\theta$ has to be drawn from a carefully specified distribution, such that the move is invariant under the target posterior distribution, i.e.,  $t_{\theta}(\bm{Z})$ follows the same distribution as the original $\bm{Z}$ under stationarity. To aid in understanding, we provide a proof in the Supplementary Material that the specified distribution of $\theta$ in Step (ii) satisfies this property. 



\subsection{Gibbs sampler for BARCW}

Under model \eqref{eq:BARCW} for BARCW, 
the Gibbs step for $\left[\bm{Z},\bm{\beta}, \bm{\alpha} \mid \Tau, \bm{W}\right]$ is very similar to that for $\left[\bm{Z},\bm{\beta}, \bm{\alpha} \mid \Tau\right]$ under model \eqref{eq:BARC} for BARC, 
with details relegated to the Supplementary Material. 
The additional step is to draw $w_j$ given all other variables. 
For $j=1,\ldots,M$, 
let $\bm{w}_{-j}$ be the weights associated with all rankers except ranker $j$. 
We draw discrete random variable $w_j$ from the following conditional posterior probability mass function:
\begin{align*}
p(w_j\mid \bm{Z},\bm{w}_{-j},\bm{\alpha},\bm{\beta},\Tau) 
&\propto   p(w_j)p(\bm{Z}\mid \bm{\alpha},\bm{\beta},\bm{w}) 
\propto  w_j^{N\over 2}
e^{-w_j
\left\| \bm{Z}_j - \bm{\alpha} - \bm{X} \bm{\beta} \right\|_2^2/2}. 
\end{align*}

\subsection{Gibbs sampler for BARCM}
Under model \eqref{eq:BARCM1}-\eqref{eq:BARCM2}, 
we first represent the parameters $\{\bm{\alpha}^{(j)}, \bm{\beta}^{(j)}\}_{j=1}^M$ by a cluster allocation vector $\bm{c} = (c_1,\ldots, c_M)$
and a set of cluster-wise parameters $\{(\bm{\alpha}^{\langle k\rangle}, \bm{\beta}^{\langle k\rangle}): k\in \{c_1,\ldots, c_M\}\}$, 
and then use an MCMC algorithm to sample $\bm{c}$, $(\bm{\alpha}^{\langle k\rangle}, \bm{\beta}^{\langle k\rangle})$'s and $\bm{Z}= (\bm{Z}_1,\ldots,\bm{Z}_M)$.

We introduce $\mathcal{R}_k(\bm{c}) = \{m: c_{m} = k, 1\leq m \leq M\}$ to denote the set of rankers that belong to cluster $k$ given cluster allocation $\bm{c}$. 
Similarly, 
let $\bm{c}_{-j}$ be the subvector of $\bm{c}$ excluding the $j$th element, 
and $\mathcal{R}_k(\bm{c}_{-j}) = \{m: c_{m} = k, m \ne j, 1\leq m \leq M\}$ 
be the set of rankers except $j$ that belong to cluster $k$.   
Due to the conjugacy between $G_0$ and the distribution of $\bm{Z}_j$'s, we can integrate out $(\bm{\alpha}^{\langle k\rangle}, \bm{\beta}^{\langle k\rangle})$'s when sampling $\bm{c}$, and the Gibbs sampling of $\bm{c}$ given $\bm{Z}$
follows from Algorithm 3 in \cite{neal2000markov}. Specifically, the Gibbs steps are as follows:
\begin{enumerate}[label = (\arabic*)]
	\item For $j=1,\dots,M$, draw $c_j$ from 
	\begin{align*}
		& \quad \ P\left(c_j = k\mid \bm{Z}, \bm{q}_{[-j]}, \Tau\right) \\
		&\propto P\left(c_j = k\mid \bm{c}_{-j}\right)\int p\left(\bm{Z}_{j}\mid \bm{\alpha}^{\langle k\rangle}, \bm{\beta}^{\langle k\rangle}\right)p\left(\bm{\alpha}^{\langle k\rangle}, \bm{\beta}^{\langle k\rangle} \mid \bm{Z}_{-j}\right) \text{d}\bm{\alpha}^{\langle k\rangle}\text{d}\bm{\beta}^{\langle k\rangle}\\
        &\propto P\left(c_j = k\mid \bm{c}_{-j}\right)\cdot\exp\left\{ -\frac{1}{2}h\big( 
        \{j\}\cup \mathcal{R}_k(\bm{c}_{-j})
        \big)+
        \frac{1}{2}
        h\big( 
        \mathcal{R}_k(\bm{c}_{-j})
        \big)
        \right\},
	\end{align*}
	where 
	$P\left(c_j \mid \bm{c}_{-j}\right)$ has the following form: 
	\begin{align*}
		P\left(c_j = k\mid \bm{c}_{-j}\right) & = \frac{|\mathcal{R}_{k}(\bm{q}_{[-j]})|}{(m-1+\gamma)}, 
		\qquad 
		\text{if } 
        k \in \{c_m: m\ne j\}
		\\
		P\left(c_j \notin \{c_m: m\ne j\} \mid \bm{c}_{-j} \right) & = \frac{\gamma}{(m-1+\gamma)},
	\end{align*}
    and $h(\cdot)$ is defined as 
	\begin{align*}
    h(\mathcal{R})
    & = 
    \sum_{m \in \mathcal{R}}
    \bm{Z}_{m}^{\top}\bm{Z}_{m}-
    \sum_{m\in \mathcal{R}} \sum_{m' \in \mathcal{R}} 
    \bm{Z}_{m}^{\top} \bm{V} 
    \left(
    \bm{\Lambda}^{-1}+ \left|\mathcal{R}\right| \bm{V}^{\top} \bm{V} \right)^{-1}
    \bm{V}^\top 
    \bm{Z}_{m'}
    \\
    & \quad \ + \log \left|\bm{\Lambda}^{-1}+ \left| \mathcal{R} \right| \bm{V}^{\top} \bm{V}\right|, 
\end{align*}
with $|\cdot|$ denoting the cardinality of a set or the determinant of a matrix.  
	
	\item For each $k\in \{c_1,\ldots,c_M\}$, we sample $[\{\bm{Z}_{j}\}_{j \in \mathcal{R}_k(\bm{c})}, \bm{\alpha}^{\langle k\rangle}, \bm{\beta}^{\langle k\rangle} \mid \Tau, \bm{c}]$ 
    using Gibbs sampling steps similar to that for 
    $[\bm{Z},\bm{\alpha},\bm{\beta} \mid \Tau]$ under the BARC model; see the Supplementary Material for details. 
\end{enumerate}

\subsection{Choice of hyperparameters}\label{sec:hyperpara}

Below  we discuss the choice of variance parameters $\sigma^2_{\alpha}$ and $\sigma^2_{\beta}$ for our Bayesian model BARC in \eqref{eq:BARC} and its extensions, 
as well as the concentration parameter $\gamma$ for the Dirichlet process in the mixture model. 
For both variance parameters, 
we impose priors following scaled inverse chi-squared distributions with parameters $(\tau^2_\alpha, \nu_\alpha)$ and $(\tau^2_\beta, \nu_\beta)$, i.e., 
$\tau^2_{\alpha} \nu_{\alpha} /\sigma^2_{\alpha}$ and $\tau^2_{\beta} \nu_{\beta} /\sigma^2_{\beta}$ follow chi squared distributions with degrees of freedom $\nu_\alpha$ and $\nu_\beta$, respectively. 
For the concentration parameter, we impose a Gamma prior with shape parameter $a_{\gamma}$ and rate parameter $b_{\gamma}$. 
The Gibbs update for $\sigma^2_\beta$ and $\sigma^2_\beta$ given $\bm{\alpha}$ and $\bm{\beta}$ under BARC and BARCW, as well as that given $\bm{\alpha}^{\langle k \rangle}$'s and $\bm{\beta}^{\langle k \rangle}$'s under BARCM and BARCMW, are straightforward and still involve sampling from scaled inverse chi-squared distributions. 
The Gibbs update for $\gamma$ involves a two-step sampling from Beta and mixture Gamma distributions as described in \citet{west1995}. 
We relegate the computation details to the Supplementary Material. 
The choice of hyperparameters $(\tau^2_{\alpha}, \nu_\alpha)$, $(\tau^2_{\beta}, \nu_{\beta})$ and $(a_{\gamma}, b_{\gamma})$ are discussed in the simulation studies. 

\section{Rank analysis via MCMC samples}\label{agg_mcmc}
Following the Bayesian computation in the previous section, we can obtain MCMC samples from the posterior distribution of $(\bm{\alpha}, \bm{\beta})$ under BARC or BARCW, and from the posterior distribution of $(\bm{\alpha}^{(j)}, \bm{\beta}^{(j)})$'s under BARCM. 
Based on the posterior samples, we can then obtain the following results from Bayesian inference.

\subsection{Aggregated ranking list and its credible interval}\label{sec:interval_rank}

Under BARC in \eqref{eq:BARC} or BARCW in \eqref{eq:BARCW}, we use the posterior means of 
$\mu_i \equiv \alpha_i+\bm{x}_{i}'\bm{\beta}$'s to generate the aggregated ranking list. 
Under BARCM in \eqref{eq:BARCM1}--\eqref{eq:BARCM2} and BARCMW, we 
use the posterior means of $M^{-1} \sum_{j=1}^{M} \mu_{i}^{(j)}  = 
M^{-1}\sum_{k} |\mathcal{R}_{k}(\bm{c})| \cdot ( \alpha_i^{\langle k\rangle}+\bm{x}_{i}'\bm{\beta}^{\langle k\rangle})$'s 
to generate the aggregated ranking list. 

Most existing rank aggregation methods seek only one aggregated rank, but ignore the uncertainty of the aggregation result. When we observe $i \succ j$ in a single aggregated ranking list, we cannot tell whether $i$ is much better than $j$ or they are close. The Bayesian inference  provides us a natural uncertainty measure for the ranking result. 
Under BARC or BARCW, 
suppose we have MCMC samples $\{\bm{\mu}^{[s]}\}_{s=1}^S$ from the posterior distribution  $p(\bm{\mu}\mid \tau_1,\cdots,\tau_M)$. 
For each sample $\bm{\mu}^{[s]}$, we calculate a ranking list
$
\rho^{[s]}=\text{rank}(\bm{\mu}^{[s]}).
$
We use $\tau^{[s]}(i)$ to denote the position of entity $i$ in ranking list $\rho^{[s]}$, and define the $(1-\alpha)$ credible interval for entity $i$'s rank as
$$
\left[ \tau_{\text{L}}(i),\tau_{\text{U}}(i)\right]=\left[\tau_{(\frac{\alpha}{2})}(i),\tau_{(1-\frac{\alpha}{2})}(i)\right],
$$
where $\tau_{(\frac{\alpha}{2})}(i)$ and $\tau_{(1-\frac{\alpha}{2})}(i)$ are the $\frac{\alpha}{2}$th and $(1-\frac{\alpha}{2})$th sample quantiles of $\{ \tau^{[s]}(i)\}_{s=1}^{S}$.
The construction of credible intervals for entities' ranks under BARCM is very similar, and thus omitted here.

\subsection{Measurement of heterogeneous rankers}

In BARCW and BARCM, as well as their combination BARCMW, we aim to learn the heterogeneity in rankers and subsequently improve 
and better 
understand 
the rank aggregation results. Both methods deliver meaningful  measures to detect heterogeneous rankers. 

In BARCW, we assume that all rankers share the same opinion and the samples from $p(\bm{w} \mid \Tau)$ measure the reliability of the input rankers. In BARCM, we assume that there exist a few groups of rankers with different opinions, despite all being reliable rankers. The MCMC samples from $p(\bm{c} \mid \Tau)$ estimate ranker clusters with different opinions. The number of clusters is determined by the number of distinct values in cluster allocation $\bm{c}$. The opinion of rankers in cluster $k$ can be aggregated by the posterior means of $\alpha_i^{\langle k\rangle}+\bm{x}_{i}'\bm{\beta}^{\langle k\rangle}$'s. We compare both methods later in simulation  and application. 

\subsection{Role of covariates in the ranking mechanism}
As discussed in Section \ref{sec:tmd_cov}, the interpretation of $\bm{\alpha}$ and $\bm{\beta}$ is difficult due to over-parameterization. 
However, noting that the $\alpha_i$'s are modeled as i.i.d Gaussian random variables with mean zero \textit{a priori}, 
the posterior distribution of $\bm{\beta}$ still provides some meaningful information about the role of covariates in the ranking mechanism. 
Intuitively, for each ranked entity $i$,  
$\bm{x}_i'\bm{\beta}$ can be viewed as the part of the evaluation score $\mu_i$ linearly explained by the covariates, and $\alpha_i$ as the corresponding residual. 
The sign and magnitude of the coefficient $\beta_k$ for the $k$th covariate indicate the positive or negative role of covariates and its strength in determining the ranking list. 
In practice, we can incorporate nonlinear transformations of original covariates to allow for more flexible role of covariates in explaining the ranking mechanism. 

\section{Simulation Studies}\label{sec:simulation}


To compare the BARC-based methods with other rank aggregation methods,  we adopt the normalized \textit{Kendall tau distance} \citep{kendall1938new} between ranking lists, which calculates the percentage of pairwise disagreements between two ranking lists. 
To measure the clustering accuracy, we adopt the Rand Index \citep{rand1971objective}, which calculates the 
percentage of pairwise clustering decisions that are correct.



\subsection{Comparison between BARC and other rank aggregation methods}\label{sec:simu_comp}

Recall that $\mathcal{U}$ is the set $\{1,\ldots,N\}$ of entities, and entity $i$ has a true score $\mu_i$. 
We generate i.i.d. covariate vectors $\bm{x}_{i} = (x_{i1}, \ldots, x_{ip})^\top$'s for the $N$ ranked entities from the multivariate Normal distribution with mean 0 and covariance $\text{Cov}(x_{is},x_{it})=\rho^{|s-t|}$ for $1\leq s,t \leq p$, 
and generate $M$ full ranking lists $\{\tau_j\}_{j=1}^M$ via the following 
model: 
\begin{align}\label{eq:model_compare}
    \tau_j=\text{rank} 
    (\bm{Z}_j), 
    \quad 
    \bm{Z}_{j} \overset{i.i.d.}{\sim} \mathcal{N}(\bm{\mu},\sigma^2\bm{I}_n), \quad (1\leq j\leq m).
\end{align}
We consider three different ways to generate the underlying true score vector $\bm{\mu}$, depending on the role of covariates. 
In Scenario 1, the true difference between entities can be linearly explained by covariates. In Scenario 2, a linear combination of covariates can partially explain the ranking. In Scenario 3, the ranking mechanism is barely correlated with the covariates. 
That is, 
\begin{enumerate}
	\item $\mu_i=\bm{x}_{i}^\top \bm{\beta}$, where  $\bm{\beta}=(3,2,-1,-0.5)^\top$, $L=4$, and $\rho=0.2$.
	\item $\mu_i=\bm{x}_{i}^\top \bm{\beta}+\|\bm{x}_{i}\|^2$, where $\bm{\beta}=(3,2,1)^\top$, $L=3$, and $\rho=0.5$.
	\item $\mu_i=\|\bm{x}_{i}\|^2$, 
	where 
	$L=4$, and $\rho=0.5$.	
\end{enumerate}

We then compare the performance of BARC with other rank aggregation methods under varying noise levels for $\sigma$ in \eqref{eq:model_compare}. 
Fixing $N=50$ and $M=10$, we tried five different values of $\sigma$ ($=1,5,10,20,40$). For each configuration, we generated 500 simulated datasets. We applied Borda Count, Markov-Chain based methods ($\text{MC}_1$, $\text{MC}_2$, $\text{MC}_3$) \citep{dwork2001rank}, 
Cross Entropy Monte Carlo based method (CEMC) \citep{lin2009integration}, 
PL model,
and our BARC model with or without covariates. 
Specifically, Borda Count uses the arithmetic means of the average ranking positions over all ranking lists, the Markov-Chain based methods are based on the stationary distributions of Markov chains whose transition matrices are constructed based on the ranking lists, 
the CEMC approach tries to find the ranking list minimizing the average distance from all ranking lists using a stochastic search method, 
and the PL model is as introduced in Section \ref{sec:PL_model}. 
A brief review of the aforementioned methods 
can be found in the Supplementary Material. 
When employing BARC and its extensions, we input standardized covariates, 
fix the degrees of freedom for scaled inverse chi-squared priors on both $\sigma^2_\alpha$ and $\sigma^2_\beta$ to be 3, i.e., $\nu_\alpha = \nu_\beta = 3$, 
and consider three choices for the scale parameters $(\tau^2_{\alpha}, \tau^2_{\beta})$: 
$(1^2, 1^2)$,  $(1^2, 10^2)$ and $(10^2, 10^2)$, denoted as $\text{BARC}_1$, $\text{BARC}_2$, $\text{BARC}_3$ respectively. 
We also consider BARC model without involving any covariates (denoted as BAR) and choose hyperparameters $\sigma_\alpha=1$, $\sigma_\beta = 10$, and $\nu_\alpha = \nu_\beta = 3$. 
Table \ref{tab:comp_other} shows 
the (scaled) Kendall tau distances between the true rank and the aggregated ranks produced by different methods, averaged over the 100 simulated datasets for each scenario and each noise level. 
Specifically, the 3rd column for the Borda Count shows the  Kendall tau distances between the estimated and true rank lists, averaged over the 100 replications, which serves as the baseline.
The remaining columns for other methods show the ratios of their average Kendall tau distances over 
the corresponding values for the Borda Count. 
First, from Table \ref{tab:comp_other}, our BARC models generally perform better than other competing methods, especially in Scenarios 1 and 2 when certain linear combination of covariates is relevant for ranking. 
Second, comparing the BARC models with and without covariates in the last four columns of Table \ref{tab:comp_other}, 
utilizing covariates can improve the precision of the rank aggregation when covariates are useful as in Scenarios 1 and 2, 
and provide little harm on precision when covariates are irrelevant as in Scenario 3. 
Third, comparing the BARC models with different prior specification in the last three columns of Table \ref{tab:comp_other}, 
we find that the results are robust to the choice of $\tau^2_{\beta}$ but sensitive to the choice of $\tau^2_{\alpha}$, 
which is related to the non-identifiability issue for $\bm{\alpha}$ and $\bm{\beta}$ as discussed in Section \ref{sec:tmd_cov}. 
Based on Table \ref{tab:comp_other}, we suggest to choose $\tau^2_\alpha = 1$ and $\tau^2_\beta = 10^2$, 
which can not only exploit the use of covariates when they are indeed relevant for ranking 
but also provide robust rank aggregation even when these covariates are irrelevant. 



\begin{table}
    \centering
    \caption{Comparison between BARC and other ranking methods. 
    The first and second columns indicate the scenario and noise level for simulating the data. 
    The 3rd column with parentheses shows the average Kendall tau distances between estimated ranking lists and corresponding true ones using Borda Count, 
    and the remaining columns without parentheses show the corresponding values for different methods but standardized by the one using Borda Count. 
    Specifically, $\text{MC}_1$--$\text{MC}_3$ denote the three forms of Markov-Chain based methods, 
    CEMC denotes the Cross Entropy Monte Carlo based method (CEMC), PL denotes the Plackett--Luce model, 
    BAR denotes our BARC mode without using any covariate, 
    and $\text{BARC}_1$--$\text{BARC}_3$ denotes our BARC model using different priors. 
    }\label{tab:comp_other}
    \resizebox{\columnwidth}{!}{%
    \begin{tabular}{cccccccccccc}
    \toprule
        Scenario & $\sigma$ & Borda &  $\text{MC}_1$ &  $\text{MC}_2$ &  $\text{MC}_3$ & CEMC & PL &  BAR & $\text{BARC}_1$ & $\text{BARC}_2$ & $\text{BARC}_3$  \\
    \midrule
    & 1 & $(0.026)$ & 1.329 & 1.083 & 1.002 & 2.749 & 4.824 & 0.999 & 0.646 & {\bf 0.643} & 0.985\\
    & 5 & $(0.124)$ & 1.285 & 1.018 & 1.009 & 1.145 & 1.098 & 0.985 & 0.653 & {\bf 0.651} & 0.972\\
 1 & 10 & $(0.216)$ & 1.498 & 1.010 & 1.001 & 1.050 & 1.068 & 0.993 & 0.705 & {\bf 0.704} & 0.983\\
   & 20 & $(0.327)$ & 1.475 & 1.004 & 0.999 & 1.024 & 1.037 & 0.996 & 0.795 & {\bf 0.793} & 0.990\\
   & 40 & $(0.408)$ & 1.193 & 1.001 & 1.000 & 1.014 & 1.000 & 0.993 & 0.884 & {\bf 0.883} & 0.990\\
    \midrule
    & 1  & $(0.028)$ & 1.280 & 1.055 & 1.000 & 2.609 & 1.093 & 0.999 & {\bf 0.981} & 0.982 & 0.988\\
    & 5 & $(0.117)$ & 1.313 & 1.016 & 1.006 & 1.120 & 1.089 & 0.989 & {\bf 0.831} & 0.832 & 0.977\\
    2 & 10  & $(0.200)$ & 1.332 & 1.013 & 1.004 & 1.056 & 1.054 & 0.993 & 0.773 & {\bf 0.772} & 0.985\\
    & 20 & $(0.299)$ & 1.406 & 1.011 & 1.000 & 1.018 & 1.042 & 0.994 & 0.813 & {\bf 0.812} & 0.988\\
    & 40 & $(0.388)$ & 1.265 & 1.002 & 1.000 & 1.005 & 1.021 & 0.996 & 0.875 & {\bf 0.874} & 0.993 \\
    \midrule
    & 1  & $(0.046)$ & 1.300 & 1.043 & 1.004 & 1.658 & 1.111 & 0.991 & 0.990 & 0.995 & {\bf 0.988} \\
    & 5  & $(0.186)$ & 1.376 & 1.011 & 1.001 & 1.054 & 1.071 & {\bf 0.991} & 1.007 & 1.007 & 0.992\\
    3 & 10 & $(0.270)$ & 1.427 & 1.008 & 1.001 & 1.033 & 1.058 & 0.996 & 1.009 & 1.011 & {\bf 0.996}\\
    & 20 & $(0.369)$ & 1.274 & 1.005 & 1.001 & 1.023 & 1.022 & 0.996 & 1.002 & 1.002 & {\bf 0.996}\\
    & 40 & $(0.421)$ & 1.196 & 1.000 & 1.000 & 1.005 & 1.028 & {\bf 0.997} & 0.997 & 0.998 & 0.997\\
    \bottomrule
    \end{tabular}%
    }
\end{table}

\begin{figure}[htb]
    \begin{subfigure}{0.32\textwidth}
        \centering
        \includegraphics[width=1\textwidth]{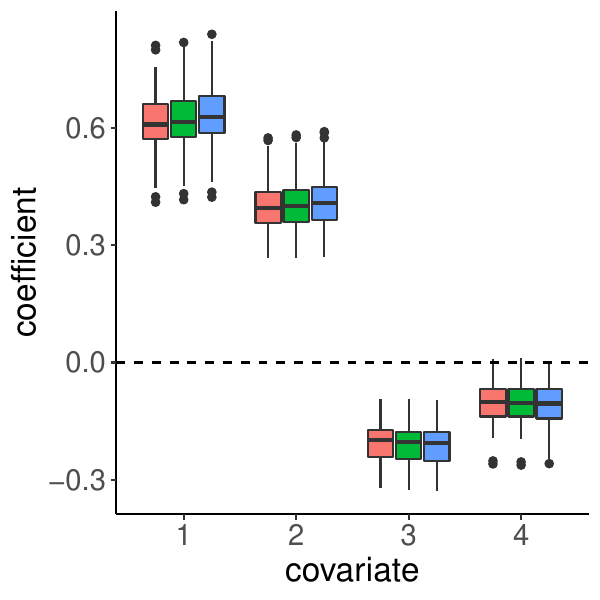}
        \caption{Scenario 1}
    \end{subfigure}%
    \begin{subfigure}{0.3\textwidth}
    \centering
        \includegraphics[width=1\textwidth]{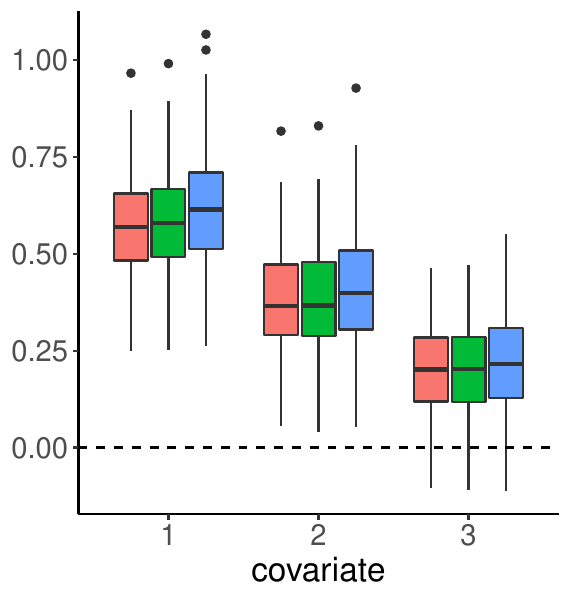}
        \caption{Scenario 2}
    \end{subfigure}%
    \begin{subfigure}{0.38\textwidth}
    \centering
        \includegraphics[width=1\textwidth]{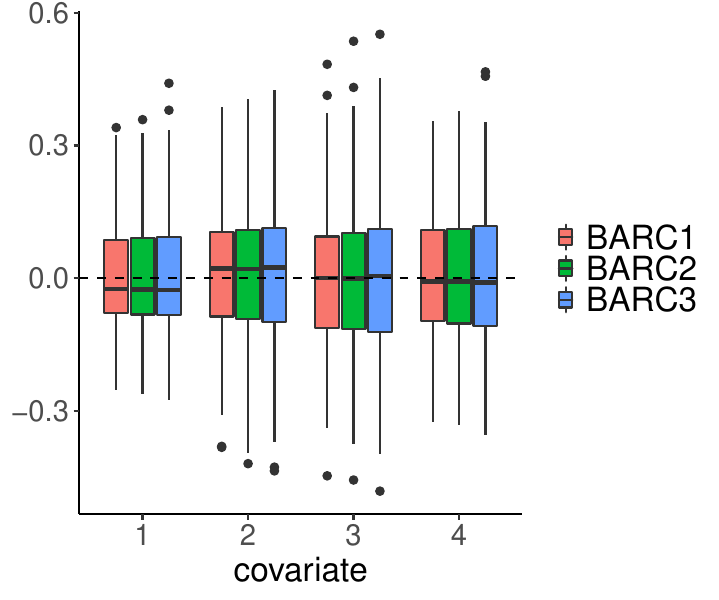}
        \caption{Scenario 3}
    \end{subfigure}
    \caption{
    Box plots of the posterior samples of the coefficients $\beta_k$'s for the standardized covariates in Scenarios 1--3 under BARC with different prior specifications. 
    }\label{fig:cov_role}
\end{figure}
We then consider the role of covariates based on our BARC models. 
Figure \ref{fig:cov_role} shows the box plots of the posterior means of the coefficients $\beta_k$'s over all 100 simulated data sets for each of the three scenarios at noise level $\sigma = 5$. 
The results from the BARC models with different prior specifications are very similar. 
Note that we fix the noise level of our BARC models at 1. It is expected that the absolute scale of our estimated coefficient may be different from the truth, even under Scenario 1 where the true score is indeed linear in the covariates. 
However, the relative magnitudes and the signs of the estimated coefficients are still informative in telling the importance of covariates for linearly explaining the ranking mechanisms, as demonstrated in Figure \ref{fig:cov_role}.

\subsection{Computational gain from parameter expansion}

\begin{figure}[htb]
    \begin{subfigure}[htbp]{0.5\textwidth}
        \centering
        \includegraphics[width=.7\textwidth]{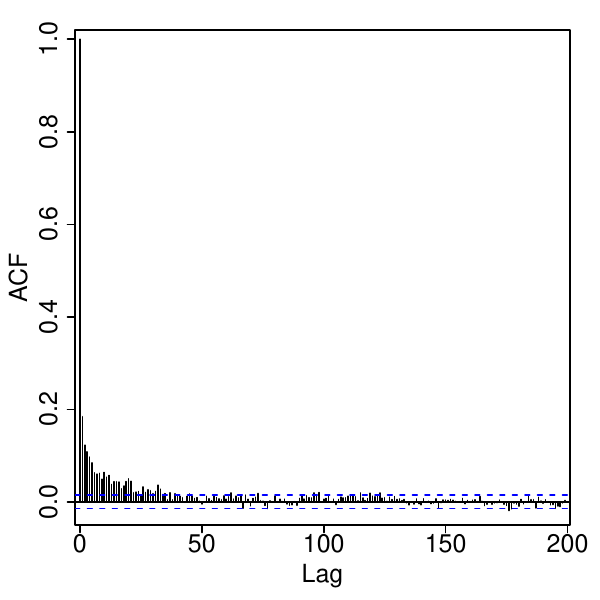}
        \caption{Parameter expanded}
    \end{subfigure}%
    \begin{subfigure}[htbp]{0.5\textwidth}
    \centering
        \includegraphics[width=.7\textwidth]{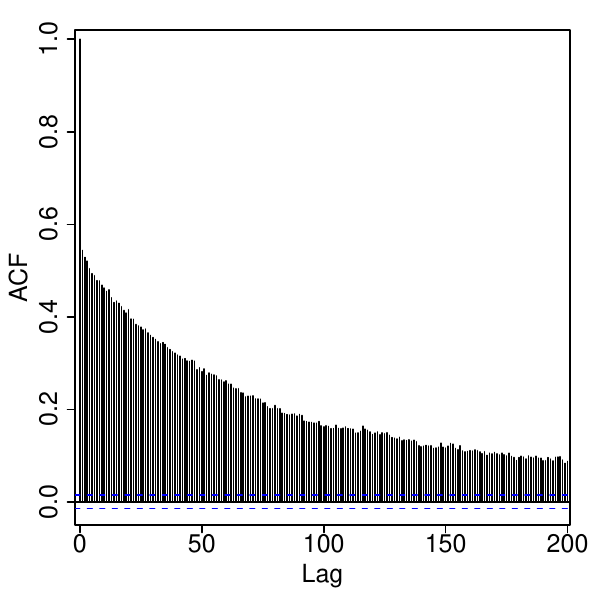}
        \caption{Regular}
    \end{subfigure}
    \caption{Auto-correlation plots of MCMC samples for $\beta_1$ in parameter expanded Gibbs sampler and regular Gibbs sampler.  
    }
    \label{fig:acf_px}
\end{figure}

Before we move on to more complex settings, we  use 
Scenario 2 with noise level $\sigma=0.1$ to demonstrate the dramatic power of parameter expansion in dealing with rank data. 
In particular, we fit the BARC model with $\sigma_\alpha$ fixed at $1$ and $\sigma_\beta$ fixed at $10$ \textit{a priori}. 
From Figure \ref{fig:acf_px}, Gibbs sampler with parameter expansion reduces the auto-correlation in MCMC samples compared to regular Gibbs sampler. We also note that  the parameter expansion step takes about the same amount of computational time as one  conditional update of a simple Gibbs sampler, which is negligible.

\begin{table}[htb]
    \centering
    \caption{Run time (in seconds) of BARC with 100 Gibbs iterations for data from Scenario 2 under various values of $(N, M, L)$}\label{tab:comp_cost}
    \resizebox{\columnwidth}{!}{%
    \begin{tabular}{cccccccccc}
    \toprule 
        Number   & \multicolumn{9}{c}{Numbers of rankers ($M$) and of covariates ($L$)}\\
        of entities  & \multicolumn{3}{c}{$M=30$} & \multicolumn{3}{c}{$M=100$} & \multicolumn{3}{c}{$M=500$}
        \\
        ($N$) & $L=5$ & $L=10$ & $L=20$ & $L=5$ & $L=10$ & $L=20$ & $L=5$ & $L=10$ & $L=20$ 
        \\
        \midrule
        $N=30$ & 1.59 & 1.36 & 1.42 & 4.89 & 5.09 & 5.37 & 24.68 & 25.49 & 25.52 \\
      $N=100$ & 5.99 & 5.83 & 6.32 & 23.20 & 23.69 & 22.95 & 105.52 & 107.96 & 107.73 \\
      $N=500$ & 96.40 & 127.03 & 143.21 & 282.68 & 283.32 & 275.85 & 1012.75 & 1003.53 &  1044.60\\
    \bottomrule
    \end{tabular}%
    }
\end{table}
Here we also briefly study the computational cost of BARC, and in particular its dependence on the number of entities $N$, number of rankers $M$ and number of covariates $L$. 
Table \ref{tab:comp_cost} shows the run times in seconds of BARC with 100 MCMC iterations for data from Scenario 2 in Section \ref{sec:simu_comp} with various values of $(N, M, L)$, using a laptop with 2.9 GHz Intel Core i9. 
From Table \ref{tab:comp_cost}, the number of covariates $L$ does not affect the run time much, 
but both the number of entities $N$ and number of rankers $M$ affect the run time considerably. 
In particular, the run time increases about linearly with $M$, and it increases faster than linearly with $N$, which implies that the number of entities matters more for the computational cost. 
The results are intuitive by noting that each Gibbs update step of BARC mainly involves sampling $NM$ truncated normal random variables for $\bm{Z}_j$'s and an $(N+L)$ multivariate normal random vector for $(\bm{\alpha}, \bm{\beta})$. 
In practice, we can parallelize the sampling for $\bm{Z}_j$'s into $M$ machines, which can reduce the run time of BARC.

\subsection{BARC with partial ranking lists}
We further explore how BARC performs for aggregating partial ranking lists, where subgroups have no overlap with each other. This is a similar situation as 
Example \ref{eg:ortho}. We simulate data from Scenario 2 with $N=80$, $M=10$ and $L=3$. We randomly divide these 80 entities into $K\,(=1,2,4,8,10,16)$ subgroups, each with size $N/K$. As $K$ increases, the pairwise comparison information decreases. 
For example, when $K=16$, we have only $5.06\%$ of all pairwise comparisons in a partial ranking list. 
\begin{table}[htb]
    \centering
    \caption{BARC for partial ranking lists. 
    The first column shows the numbers of non-overlapping subgroups of equal sizes, under which we only observe ranks within each subgroup.
    The second to fifth columns show the average Kendall tau distances between estimated ranking lists and corresponding true ones using different forms of our BARC model. 
    Specifically, BAR denotes our BARC mode without using any covariate, 
    and $\text{BARC}_1$--$\text{BARC}_3$ denotes our BARC model using different priors.
    }
    \label{tab:partial}
    \begin{tabular}{ccccc}
    \toprule
       Number of subgroups  & BAR & $\text{BARC}_1$ & $\text{BARC}_2$ & $\text{BARC}_3$  \\
       \midrule
        1 & 0.120 & 0.101 & {\bf 0.101} & 0.117 \\
        2 & 0.127 & 0.101 & {\bf 0.100} & 0.126 \\
        4 & 0.136 & {\bf 0.103} & 0.103 & 0.135 \\
        8 & 0.162 & {\bf 0.112} & 0.112 & 0.144\\
        10 & 0.173 & {\bf 0.111} & 0.111 & 0.146 \\
        16& 0.203 & {\bf 0.117} & 0.117 & 0.154 \\
        \bottomrule
    \end{tabular}
\end{table}
Table \ref{tab:partial}
displays the Kendall tau distances between the true and the aggregated ranking lists inferred by BARC models in different cases. 
Specifically, we consider four Bayesian models BAR, $\text{BARC}_1$, $\text{BARC}_2$ and $\text{BARC}_3$, whose models and priors are defined the same as in Section \ref{sec:simu_comp}. 
Table \ref{tab:partial} shows that
BARC is quite robust with respect to partial ranking lists when the unobserved pairwise comparisons are missing completely at random and the input ranking lists have moderate dependence on the available covariates, 
in the sense that the precision of aggregated ranking lists is relatively stable when the partial ranking lists become more and more incomplete, as demonstrated by $\text{BARC}_1$ and $\text{BARC}_2$. 
In contrast, the BARC method without using covariates, denoted by BAR in Table \ref{tab:partial}, is susceptible to missing information in the partial lists. 
Specifically, when $K=16$, the average Kendall tau distance between true and aggregated ranking lists using BAR increases by about $70\%$, while that using $\text{BARC}_1$ and $\text{BARC}_2$ increases by about $15\%$.  



\subsection{BARCW for rankers with varying qualities}\label{sec:simu_quality}

\begin{figure}[htbp]
    \begin{subfigure}[htbp]{0.33\textwidth}
        \centering
        \includegraphics[width=1\textwidth]{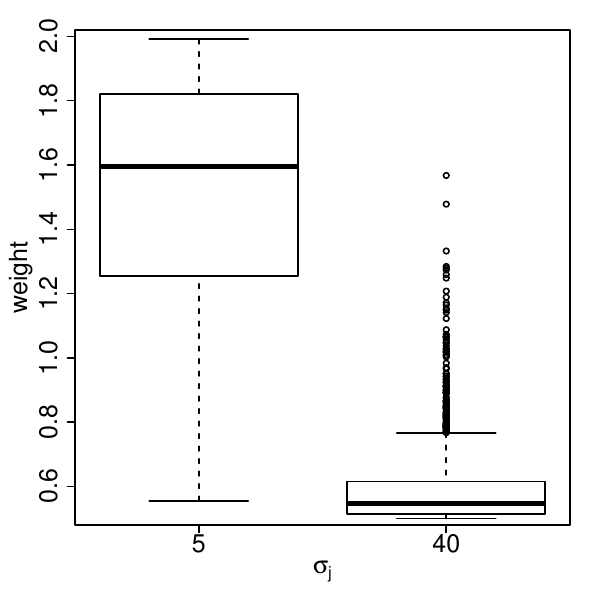}
        \caption{Section \ref{sec:simu_quality}}
    \end{subfigure}%
    \begin{subfigure}[htbp]{0.33\textwidth}
    \centering
        \includegraphics[width=1\textwidth]{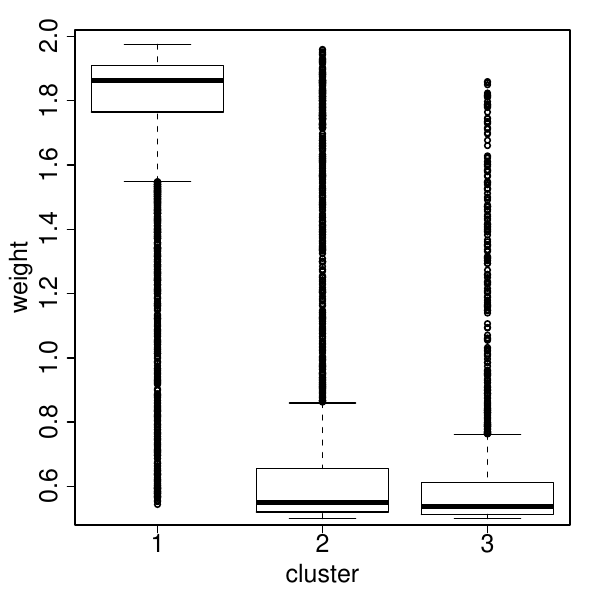}
        \caption{Section \ref{sec:simu_barcm}}
    \end{subfigure}%
    \begin{subfigure}[htbp]{0.33\textwidth}
    \centering
        \includegraphics[width=1\textwidth]{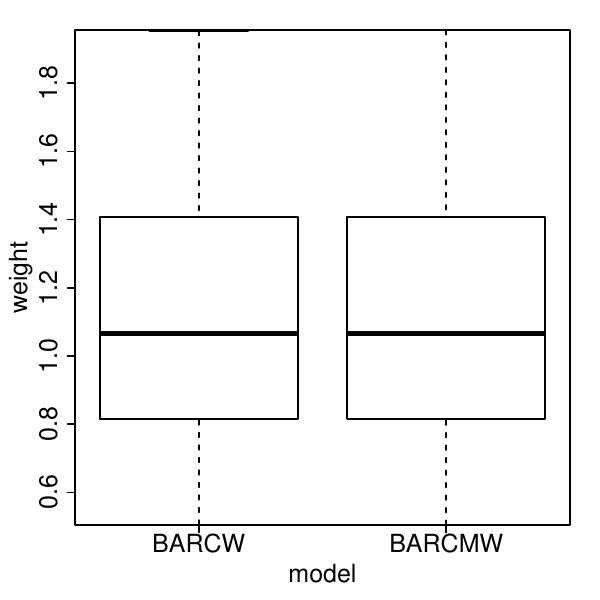}
        \caption{Section \ref{sec:homogeneous}}
    \end{subfigure}
    \caption{
    Box plots of the posterior means of weights for certain subgroup of rankers. 
    Specifically, 
    (a) shows the box plots for rankers with noise levels $\sigma_j = 5$ and $\sigma_j = 40$, respectively, over all simulations from Scenarios 1--3 in  Section \ref{sec:simu_quality}. 
    (b) shows the box plots for rankers in different clusters over all simulations from Scenario I in Section \ref{sec:simu_barcm}. 
    (c) shows the box plots for all rankers over all simulations in Section \ref{sec:homogeneous} using models BARCW and BARCMW, respectively. 
    }
    \label{fig:boxplot_weight}
\end{figure}

We investigate the setting where the rankers have consistent opinions but various qualities. 
The data are simulated from Scenarios 1--3 in Section \ref{sec:simu_comp} with $N=80$ and $M=10$, except that the noise level $\sigma$ in \eqref{eq:model_compare} is allowed to be ranker-specific. 
Specifically, half of the rankers have noise level $\sigma_j = 5$ and the remaining half have noise level $\sigma_j = 40$, which represent high-quality and low-quality rankers respectively. 
\begin{table}[htb]
    \centering
    \caption{Comparison between BARC and other ranking methods when there are rankers of varying qualities.
    The first column shows the scenario for simulating the data. 
    The second column with parentheses shows the average Kendall tau distances between estimated ranking lists and corresponding true ones using Borda Count, 
    and the remaining columns without parentheses show the corresponding values for different methods but standardized by the one using Borda Count. 
    Specifically, $\text{MC}_1$--$\text{MC}_3$ denote the three forms of Markov-Chain based methods, 
    CEMC denotes the Cross Entropy Monte Carlo based method (CEMC), PL denotes the Plackett--Luce model, 
    BARC denotes our BARC mode,
    and BARCW denotes our BARCW model. 
    }
    \begin{tabular}{ccccccccc}
    \toprule
        Scenario  & Borda &  $\text{MC}_1$ &  $\text{MC}_2$ &  $\text{MC}_3$ & CEMC & PL &  BARC & BARCW \\
    \midrule
    1 & $(0.214)$ & 1.922 & 0.972 & 1.009 & 1.024 & 1.041 & 0.676 & {\bf 0.579} \\
    2 & $(0.200)$ & 1.866 & 0.950 & 1.015 & 1.007 & 1.052 & 0.753 & {\bf 0.651} \\
    3 & $(0.275)$ & 1.625 & 0.989 & 1.005 & 1.008 & 1.044 & 0.998 & {\bf 0.943} \\
    \bottomrule
    \end{tabular}%
\end{table}
Table \ref{tab:comp_other} shows 
the (scaled) Kendall tau distances between the true rank and the aggregated ranks produced by different methods, averaged over the 100 simulated datasets for each scenario, 
where the values for all methods other than Borda Count are standardized by the corresponding values for Borda Count. 
Moreover, we use the same prior specification for BARC and BARCW models, i.e., $\sigma_\alpha = 1$, $\sigma_\beta = 10$ and $\nu_\alpha = \nu_\beta = 3$. 
From Table \ref{tab:comp_other}, 
BARC shows more precise rank aggregation than other methods under comparison, especially when the covariates can linearly explain the underlying true scores for ranking. 
By exploring the varying qualities of the rankers, BARCW further improves BARC, and the improvement increases with the role of covariates for linearly explaining the ranking mechanisms. 
Figure \ref{fig:boxplot_weight}(a) shows the box plots of the posterior means of the weights for high-quality and low-quality rankers, separately. 
From Figure \ref{fig:boxplot_weight}(a), 
the weights for rankers with different qualities are well separated, and thus 
BARCW is able to identify rankers of different qualities.

\subsection{BARCM for rankers with heterogeneous opinions}\label{sec:simu_barcm}
In many applications, there can be multiple groups of rankers with different opinions, despite all being reliable rankers. The Dirichlet process mixture model \eqref{eq:BARCM1}-\eqref{eq:BARCM2} can be used to determine the total number of clusters and to cluster the rankers. 
Here, we use simulations to test the sensitivity of BARCM to hyperparameter settings in the Dirichlet process prior. 
In particular, following the discussion in Section \ref{sec:hyperpara}, we fix the degrees of freedom $\nu_\alpha$ and $\nu_\beta$ at 3, 
and study the sensitivity of  BARCM to the scale hyperparameters $\tau^2_\alpha$ and $\tau^2_\beta$ for the variances $\sigma^2_\alpha$ and $\sigma^2_\beta$ of latent variables and hyperparameters $a_\gamma$ and $b_\gamma$ for the concentration parameter $\gamma$ of the Dirichlet process.  
For the variances, we consider two choices of $(\tau_\alpha, \tau_\beta)$: $(1, 10)$ and $(10, 10)$. 
For the concentration parameter, we consider two choices of $(a_\gamma, b_\gamma)$ 
suggested by 
\citet{west1995} and \citet{fruhwirth2019}: $(2, 4)$ and $(1, 20)$, 
where the latter implies a smaller number of clusters \textit{a priori}; 
see \citet{fruhwirth2019} for more detailed discussion. 
The combination of these choices results in four prior specifications for the BARCM model, 
and we denote them by $\text{BARCM}_1$, $\text{BARCM}_2$, $\text{BARCM}_3$ and $\text{BARCM}_4$ respectively, as shown in Table \ref{tab:BARCM}.

We consider two simulation scenarios under the BARC model as in \eqref{eq:BARC} but with three mixture components. 
For each scenario, 
we have $N$ entities evenly divided into $G$ non-overlapping subgroups, $L$ covariates for each entity, and $M$ rankers who rank only entities within the same subgroup. 
The categories of rankers are generated from a Multinomial distribution with probabilities $(0.5, 0.3, 0.2)$. 
The covariates $\bm{x}_i$'s are generated from a multivariate Normal distribution with mean zero and pairwise covariances $Cov(x_{is},x_{it})=(0.2)^{|s-t|}$ for $1\le s, t\le L$, 
the coefficients are generated from $\bm{\alpha}^{\langle k \rangle}\overset{i.i.d.}{\sim}\mathcal{N}(0, 4\bm{I}_N)$ and $\bm{\beta}^{\langle k \rangle}\overset{i.i.d.}{\sim} \mathcal{N}(0,\bm{I}_L)$ for $1\le k\le 3$ , 
and the noise level is fixed at $1$.
In Scenario I, 
we choose $N = 20$, $G=1$, $L=3$ and $M=100$, i.e., each ranker provides a full ranking list for all the entities. 
In Scenario II, we choose $N=108$, $G=9$, $L=11$ and $M = 69$, i.e., each ranker provides only a partial ranking list comparing units within the same subgroup of $N/G = 12$ units. 
Scenario II mimics the dataset in Example \ref{eg:ortho}. 

Table \ref{tab:BARCM} shows the accuracy (measured by the Rand Index) of the {\it maximum a posteriori} (MAP) estimate of the clustering indicators  and the posterior expected number of clusters from the BARCM model with different hyperparameters, averaged over 100 simulated datasets. 
From Table \ref{tab:BARCM}, 
the results are relatively robust with respect to different choices of priors, 
although a large value of $\tau_{\alpha}$ leads to a slight overestimation of the number of clusters. 
Intuitively, this may be due to the fact that a larger value of $\tau_{\alpha}$ implies a larger signal noise ratio, thus requesting 
more consistent rankings among rankers in the same cluster and encouraging more clusters of rankers.  
\begin{table}[htb]
    \centering
    \caption{Accuracy of the clustering assignments and posterior expected number of clusters
    under  BARCM with different prior specifications. 
    }
    \label{tab:BARCM}
    \resizebox{\columnwidth}{!}{%
    \begin{tabular}{cccccc}
    \toprule
    Scenario &  & $\text{BARCM}_1$ & $\text{BARCM}_2$ & $\text{BARCM}_3$ & $\text{BARCM}_4$ \\
    \midrule
     & $(\tau_\alpha, \tau_\beta, a_\gamma, b_\gamma)$ & (1, 10, 2, 4) & (10, 10, 2, 4) & (1, 10, 1, 20) & (10, 10, 1, 20)
    \\
    \midrule
    \multirow{2}{*}{I} & Clustering accuracy & 1.000 & 0.999 & 1.000 & 0.999 \\
        & Expected \# of clusters  & 3.000 & 3.007 & 3.001 & 2.991 \\
    \midrule
    \multirow{2}{*}{II}    & Clustering accuracy & 0.999 & 0.988 & 0.999 & 0.987\\
        & Expected \# of clusters  & 3.010 & 3.300 & 3.010 & 3.312 \\
    \bottomrule
    \end{tabular}%
    }
\end{table}


Here we also explore the performance of BARCW when there are indeed mixture subgroups of rankers with different ranking opinions, i.e., using a weighting strategy to construct a ``consensus". 
We fit the BARCW model under the prior that $\tau_\alpha = 1$, $\tau_\beta = 10$ and $\nu_\alpha = \nu_\beta = 1$. 
Figure \ref{fig:boxplot_weight}(b) shows the box plots of the posterior means of weights for rankers in different clusters over all 100 simulated data sets from Scenario I, which demonstrates 
that the majority opinions are up-weighted by BARCW, while the other opinions are down-weighted. 
This is intuitive and expected since BARCW assumes that all rankers share the same opinion. 
As a result, BARCW reinforces the majority's opinion in rank aggregation. 
By studying rankers' heterogeneity using either BARCW or BARCM, we can better understand our ranking data even if we seek only one aggregated ranking list.

\subsection{Robustness of BARCM and BARCW under homogeneous setting}\label{sec:homogeneous}

In contrast to the simulation with heterogeneous rankers, 
we also simulated the BARC model under the homogeneous setting to verify the robustness of BARCM and BARCW, as well as BARCMW. 
The simulation is the same as Scenario I in Section \ref{sec:simu_barcm} except that all rankers are from one component with equal qualities. 
We fit the BARCW, BARCM and BARCMW models with hyperparameters $\tau_\alpha = 1$, $\tau_\beta = 10$, 
$\nu_\alpha = \nu_\beta = 3$, $a_\gamma = 2$ and $b_\gamma = 4$. 
The clustering accuracy for BARCM and BARCMW averaged over all 100 simulated datasets are respectively, 0.9996 and 1. 
Thus, both models classify the rankers into one cluster, i.e., the rankers have consist opinions. 
Figure \ref{fig:boxplot_weight}(c) shows the box plots of the posterior means of weights for all rankers from BARCW and BARCMW over all 100 simulated datasets. 
From Figure \ref{fig:boxplot_weight}(c), all rankers have similar qualities, which is consistent with the true data generation models. 

\section{Analyses of the Two Real Data Sets}\label{sec:realdata}

Below we will analyze the two applications in Examples \ref{eg:nfl} and \ref{eg:ortho} using our Bayesian models. 
Based on the simulation studies in Section \ref{sec:simulation}, 
we  set the hyper-parameters for the Bayesian models to be $\tau_\alpha = 1$, $\tau_\beta = 10$, 
$\nu_\alpha = \nu_\beta = 3$, $a_\gamma = 2$ and $b_\gamma = 4$. 

\subsection{Aggregating NFL quarterback rankings}\label{sec:NFL}
\begin{figure}[htbp]
    \begin{subfigure}[htbp]{0.5\textwidth}
        \centering
        \includegraphics[width=.78\textwidth]{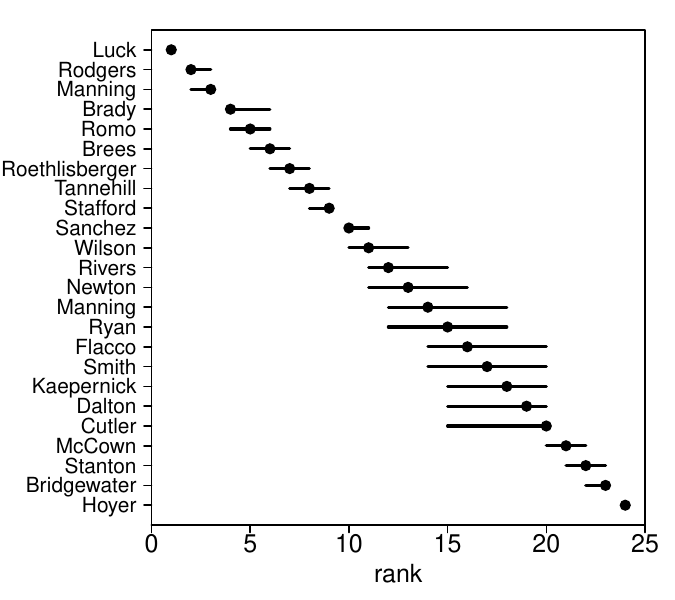}
        \caption{Ranked position}
    \end{subfigure}%
    \begin{subfigure}[htbp]{0.5\textwidth}
    \centering
        \includegraphics[width=.7\textwidth]{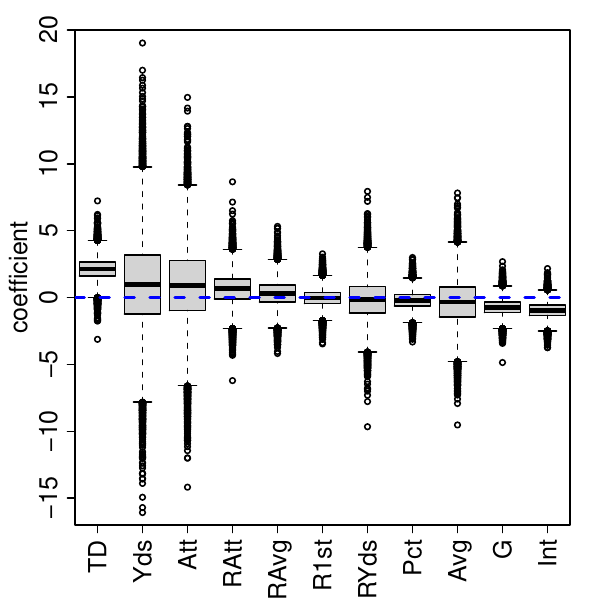}
        \caption{Covariates' coefficients
        }
    \end{subfigure}
    \caption{
    (a) shows the credible intervals of the ranked position of the NFL quarterbacks under BARCW, 
    and (b) shows the corresponding box plots of the posterior samples of the coefficients $\beta_l$'s for the standardized covariates. 
    }\label{fig:nfl_barcw}
\end{figure}

Ranking  NFL quarterbacks is a classic case where experts' ranking schemes are clearly related to some performance statistics of the players in their games. Information in Tables \ref{tab:NFL_rank_data} and \ref{tab:NFL_covariate} enables us to generate aggregated lists using both rank data and the covariate information, as shown in Table \ref{tab:nfl_agg}. 
For quarterbacks at the top and bottom of the list, these methods mostly agree with each other. 
\begin{table}[htb]
    \centering
    \caption{NFL rank aggregation using different methods. 
    The first column shows the players' names, and the remaining columns show their ranked positions using different methods. 
    BARC, BARCW, BARCM and BARCMW denote our Bayesian models for rank data, 
    Borda denotes Borda Count, and $\text{MC}_3$ denotes the Markov-Chain based method.
    }\label{tab:nfl_agg}
    \resizebox{\columnwidth}{!}{%
    \begin{tabular}{p{3cm}p{1.3cm}p{1.3cm}p{1.3 cm}p{1.3cm}p{1.3cm}p{1.3cm}}
\toprule
Player &   BARC & BARCW & BARCM & BARCMW & Borda & $\text{MC}_3$\\
\midrule
Andrew Luck         & 1&    1&   1&   1&   1 &  1 \\
Aaron Rodgers       & 2&    2&   2&   2&   2 &  2\\
Peyton Manning      & 3&    3&   3&   3&   3 &  3\\
Tom Brady           & 4&    4&   4&   4&   4 &  4\\
Tony Romo           & 5&    5&   5&   5&   5 &  5\\
Drew Brees          & 6&    6&   6&   6&   6 &  6\\
Ben Roethlisberger  & 7&    7&   7&   7&   7 &  7\\
Ryan Tannehill      & 8&    8&   8&   8&   8 &  8\\
Matthew Stafford    & 9&    9&   9&   9&    9 &  9\\
Mark Sanchez        & 10&  10&  10&  10&  10 &  10\\
Russell Wilson      & 11&  11&  11&  11&  11 &  11\\
Philip Rivers       & 12&  12&  12&  12&  12 &  12\\
Cam Newton          & 13&  13&  13&  13&  13 &  13\\
Eli Manning         & 14&  14&  14&  14&  14 &  14\\
Matt Ryan           & 15&  15&  15&  15&  15 &  15\\
Joe Flacco          & 19&  16&  19&  16&  19 &  19\\
Alex Smith          & 17&  17&  17&  17&  17 &  17\\
Colin Kaepernick    & 16&  18&  16&  18&  16 &  16\\
Andy Dalton         & 20&  19&  20&  19&  20 &  20\\
Jay Cutler          & 18&  20&  18&  20&  18 &  18\\
Josh McCown         & 21&  21&  21&  21&  21 &  21\\
Drew Stanton        & 22&  22&  22&  22&  22 &  22\\
Teddy Bridgewater   & 23&  23&  23&  23&  23 &  23\\
Brian Hoyer         & 24&  24&  24&  24&  24 &  24\\
\bottomrule
    \end{tabular}%
    }
\end{table}
Besides only looking at the aggregated ranking lists, as suggested in Section \ref{sec:interval_rank}, 
it is important to investigate the uncertainty in rank aggregation, which can help mitigate and explain the discrepancy across different methods. 
Using BARCW as an example, 
Figure \ref{fig:nfl_barcw}(a) shows the $95\%$ credible intervals for all quarterbacks' ranked positions under BARCW. We can see that the interval width is large for mediocre quarterbacks, which is exactly where a majority of discrepancies occurred among different rankers and different rank aggregation methods. The interval estimates of aggregated ranks can separate several elite quarterbacks from the others. 
In practice, this may suggest an aggregated ranking list with ties or a bucket order for several subgroups of entities, instead of a full ranking list with considerable uncertainties; 
see, e.g.,  \citet{medianconstrainedbucket2019}, \citet{kenkre2011discovering} and \citet{gionis2006algorithms} for related discussions.

\begin{figure}[ht]
	\begin{subfigure}[htbp]{0.5\textwidth}
		\centering
		\includegraphics[width=.7\textwidth]{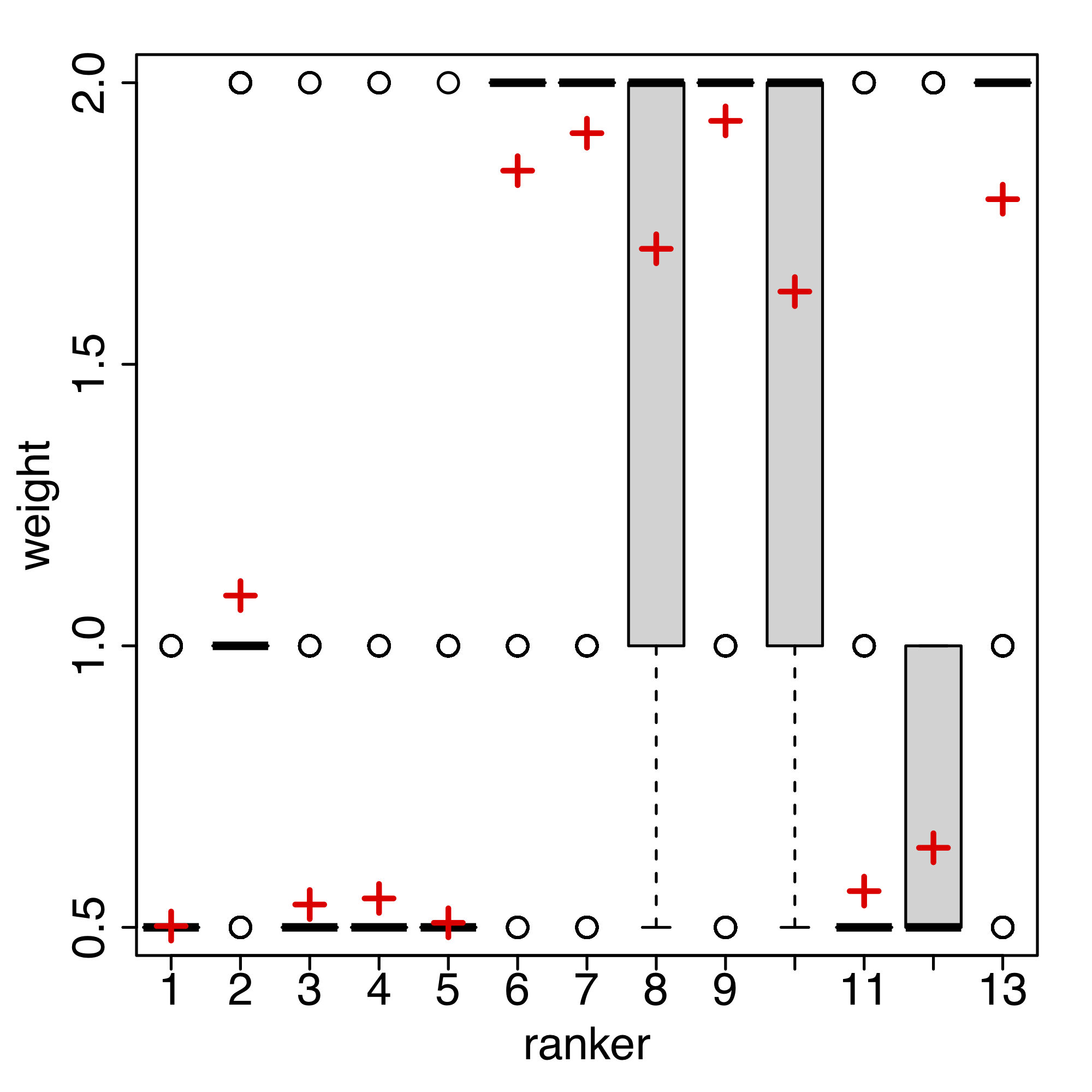}
		\caption{Box plot of weights}
	\end{subfigure}%
	\begin{subfigure}[htbp]{0.5\textwidth}
		\centering
		\includegraphics[width=.7\textwidth]{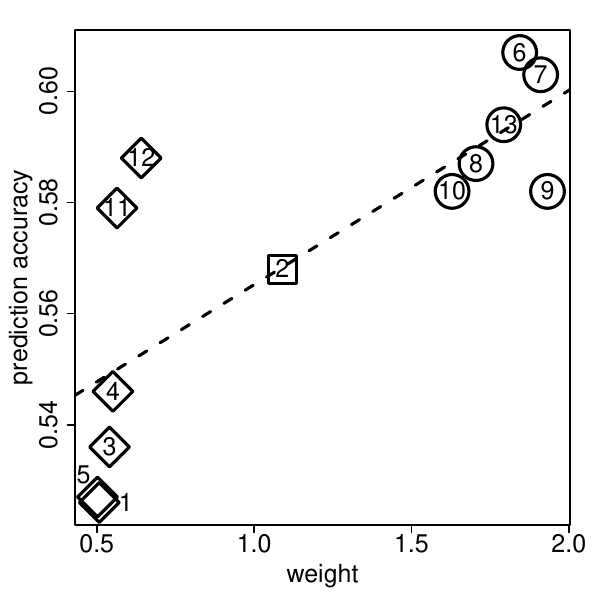}
		\caption{Accuracy versus weight}
	\end{subfigure}
	\caption{
	(a) shows the box plots of the posterior samples of the weights for all rankers under BARCW, 
	and (b) plots the prediction accuracy against the posterior mean of the weight for all rankers.
	}\label{fig:NFL_weight}
\end{figure}

All methods except BARCW, BARCM and BARCMW assume equal reliability for all rankers. 
After analyzing the data using BARCW, we show in 
Figure \ref{fig:NFL_weight}(a) the box plots as well as the means of the posterior samples of the weights for all rankers.
Out of the 13 rankers, six are inferred to have significantly higher quality than the others with a majority of their posterior samples of weights being greater than or equal to 1. 
The second ranker seems to have medium quality with weight close to 1, while the remaining rankers all have weights close to 0.5. 
We further validate our weight 
estimation using the prediction accuracy of the experts at the end of the season. 
Figure \ref{fig:NFL_weight}(b) plots this prediction accuracy against the posterior mean weight  of each ranker resulting from BARCW, which shows that
rankers with higher weights predicts more accurately
on average, and the correlation between these two measures is  quite high at 0.784.

Under either BARCM or BARCMW,  the 13 rankers are clustered into subgroups with different ranking opinions. 
To avoid the impact of multimodal posterior distributions,  
we run 100 MCMC chains with different random initial starts under either BARCM or BARCMW, 
and then choose the {\it maximum a posteriori} (MAP) estimate 
(i.e., the one with the highest joint posterior density). 
The resulting MAP estimates of clustering under both BARCM and BARCMW suggest that all the 13 experts 
belong to the same cluster, strongly suggesting that
these experts share the same ranking opinion but have different qualities. 
Consequently, 
BARCW seems to be the most appropriate model for this application. 

We further investigate the role of covariates in ranking these players. 
Figure \ref{fig:nfl_barcw}(b) shows the posterior means and $95\%$ posterior credible intervals for the coefficients of the eleven standardized covariates listed in Table \ref{tab:NFL_covariate}.
TD and Int, which stand for percentage of touchdowns and interceptions thrown when attempting to pass, are the most significant covariates; touchdowns have a positive effect, while interceptions have a negative effect. 
Based on the football common sense, touchdowns and interceptions can directly impact the result of a game.

\subsection{Aggregating orthodontics data}

As mentioned in Section \ref{sec:intro}, the orthodontics data set contains 69 partial ranking lists for each of the 9 groups of the orthodontic cases. With ranking lists produced by a group of high-profile specialists, the rank aggregation problem emerges because the average perception of experienced orthodontists is regarded as the cornerstone of systems for the evaluation of orthodontic treatment outcome \citep{liu2012consistency, song2014reliability, song2015validation}. The covariates for these cases are objective assessments on their teeth. It is quite difficult to aggregate ranking lists of many non-overlapping subgroups, as covariates are the only source of information available in bridging different groups. In addition, Table \ref{tab:ortho_rank_data} shows that the rankers do have significantly different opinions. 

\begin{figure}[htb]
	\centering
	\includegraphics[width=1\textwidth]{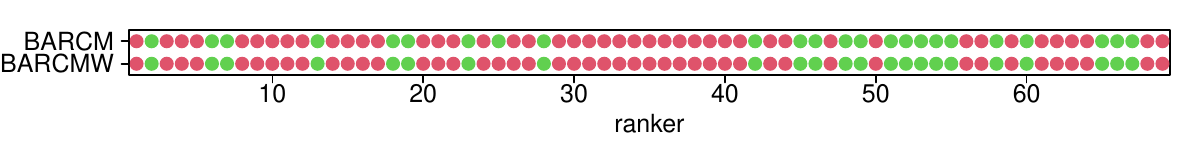}
	\caption{
	The MAP estimates of the clustering of the rankers under BARCM and BARCMW. 
	}\label{fig:Ortho_BARCM}
\end{figure}

Previously, \cite{liu2012consistency} and \cite{song2014reliability} assessed the reliability and the overall consistency of these experienced orthodontists through simple statistics including Spearman's correlation among these highly incomplete ranking lists within each subgroup of cases.
To gain a deeper understanding of these ranking lists, we first study the heterogeneity among rankers using BARCM and BARCMW.
To avoid the impact of multimodal posterior distributions, similar to Section \ref{sec:NFL}, 
we run 100 MCMC chains with random initial starts under either BARCM or BARCMW, and choose the one that gives rise to the MAP estimate.
Figure \ref{fig:Ortho_BARCM} shows the MAP estimates of the clusters, demonstrating that the 69 experts are clustered into 2 subgroups  of sizes $(45, 24)$, and the clustering of rankers using BARCM and BARCMW are quite consistent,
where only one expert is clustered differently. 
About $45/69 = 65.2\%$ of the experts share consistent opinions about the ranking of the 108 patients, while the remaining experts rank the patients in a different way. 
This implies that most discrepancy among the experts for ranking the patients should not be explained by the quality variations of the experts, but are attributable to their different opinions. 

\begin{figure}[htb]
	\begin{subfigure}[htbp]{0.45\textwidth}
		\centering
		\includegraphics[width=1\textwidth, height=0.2\textheight]{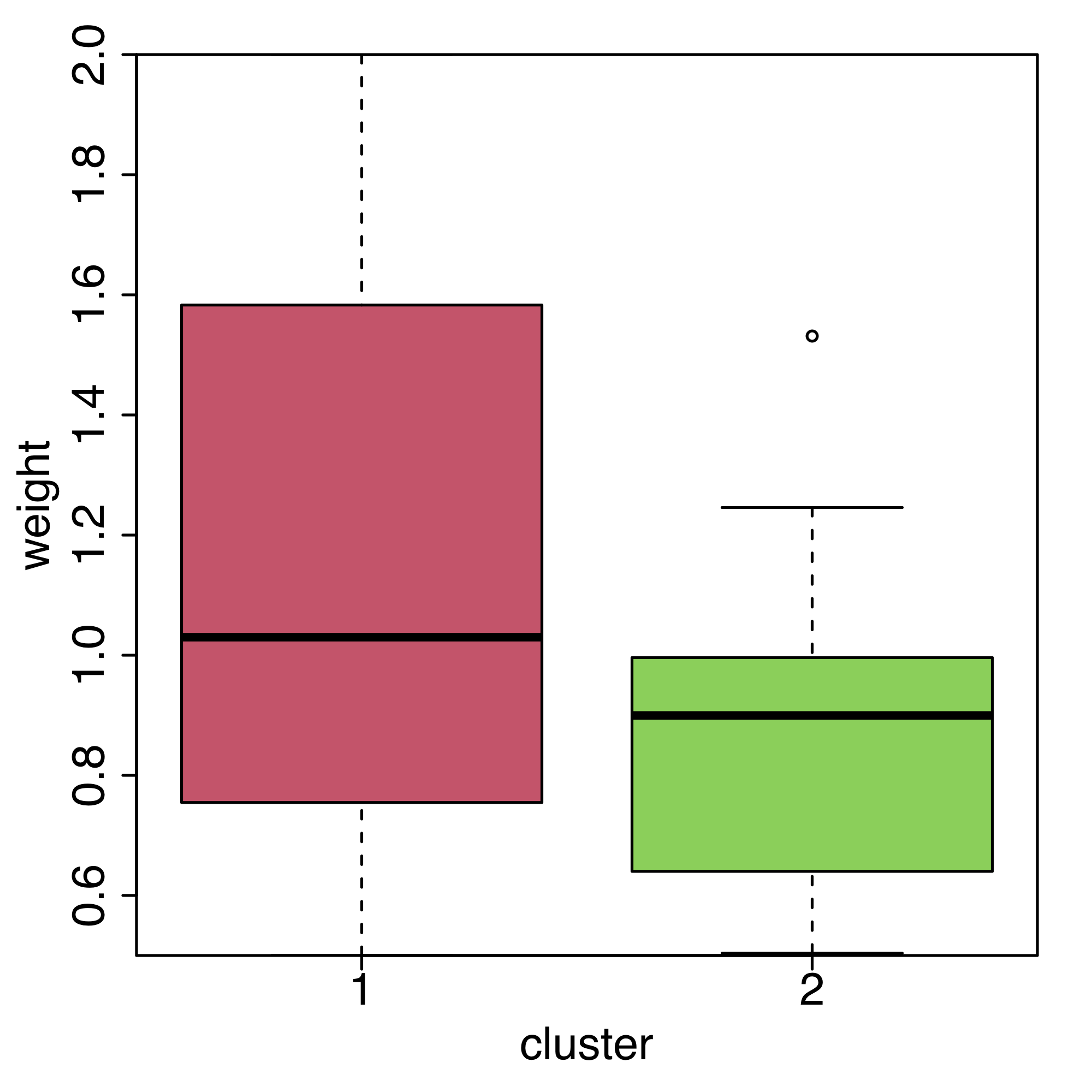}
		\caption{Ranker's weight}
	\end{subfigure}%
	\begin{subfigure}[htbp]{0.45\textwidth}
		\centering
		\includegraphics[width=1\textwidth,height=0.2\textheight]{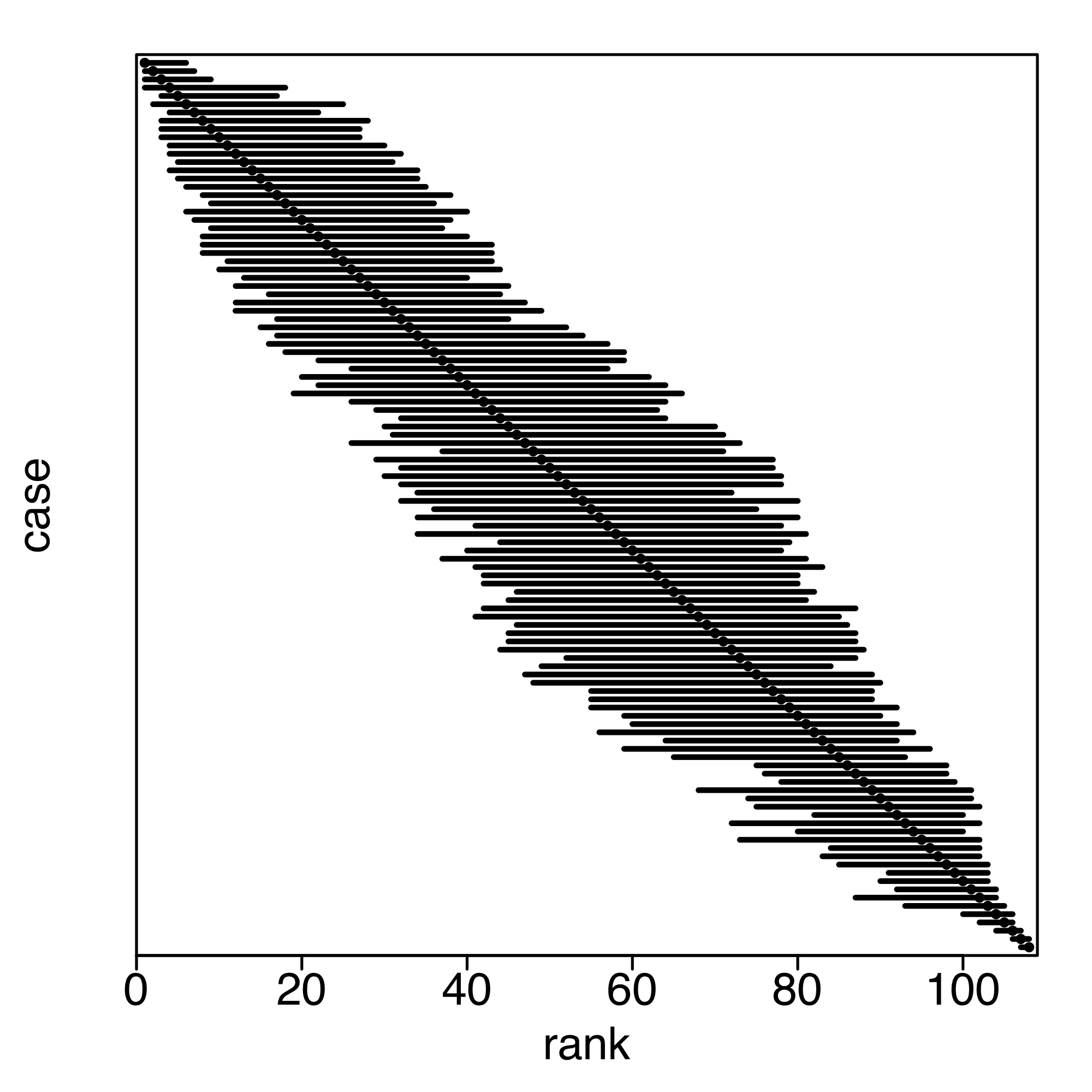}
		\caption{Rank interval}
	\end{subfigure}
	\caption{
	(a) shows the box plot of the posterior means of weights for rankers in different clusters estimated by BARCM. 
	(b) shows the $95\%$ credible intervals of the ranked positions of the cases under BARCM. 
	}\label{fig:BARCM_weight_rank}
\end{figure}

Figure \ref{fig:BARCM_weight_rank}(a) shows the box plot of rankers' weights resulting from BARCW by their estimated clusters from BARCM. 
 A majority of rankers in the larger cluster are labeled as reliable rankers, 
and most rankers in the smaller cluster are labeled as mediocre or low-quality rankers. 
This result is similar to our simulation results in Section \ref{sec:simu_barcm}, i.e.,
in order to form a ``consensus",
BARCW  down-weights the minority opinions when heterogeneous opinions exist. 

We then study rank aggregation using our Bayesian models. The key to aggregating these nine non-overlapping groups of patients is to figure out the rank of patients' orthodontics conditions using, but not overly relying on, the covariates.
Table \ref{tab:ortho_rank} shows the top and bottom cases in aggregated ranking lists using different models, as well as aggregated ranking lists for each cluster under BARCM.  
Recall that BARCM aggregates opinions of the whole sample by averaging over all clusters with their corresponding proportions. The results from BARCW and BARCM are quite consistent with each other although they employ different assumptions. The Kendall tau distance between these two aggregated lists is 0.043. 
\begin{table}
	\caption{
	The five cases that are considered to have the best and worst conditions based on rank aggregation. 
	The first column denotes the ranked position. 
	The second to fifth columns show the top and bottom five cases in the aggregated ranking lists from our Bayesian models for rank data. 
	The last two columns show the top and bottom five cases in the aggregated ranking lists for the two clusters of rankers estimated from BARCM. 
	}\label{tab:ortho_rank}
	\begin{tabular}{ccccccc}
		\toprule
		& BARC & BARCW & BARCM & BARCWM & Cluster 1 & Cluster 2 \\
		\midrule
		1 & G7 & G7 & G7 & G7 & H2 & G7 \\ 
		2 & H2 & E2 & E2 & E2 & G7 & E2 \\ 
		3 & E2 & H2 & H2 & H2 & E2 & A1 \\
		4 & H3 & H3 & F8 & F8 & F8 & E10\\
		5 & H4 & H4 & H3 & H3 & H3 & E1\\
		\midrule
		104 & E6 & D11& D11& D11& E6 & D11 \\ 
		105 & D11& E6 & E6 & E6 & D11& E6  \\
		106 & F10& F10& F10& F10& F10& F10 \\
		107 & H5 & H5 & H5 & F4 & F4 & H5  \\
		108 & F4 & F4 & F4 & H5 & H5 & F4  \\
		\bottomrule
	\end{tabular}
\end{table}
Figure \ref{fig:BARCM_weight_rank}(b) shows the $95\%$ credible intervals for the ranked positions of the 108 cases, demonstrating that
there is a substantial amount of uncertainty in the aggregated ranking list, especially around the middle of the ranking list. 

\begin{figure}[htb]
	\centering
	\begin{subfigure}{.5\textwidth}
		\centering
		\includegraphics[width=0.9\linewidth]{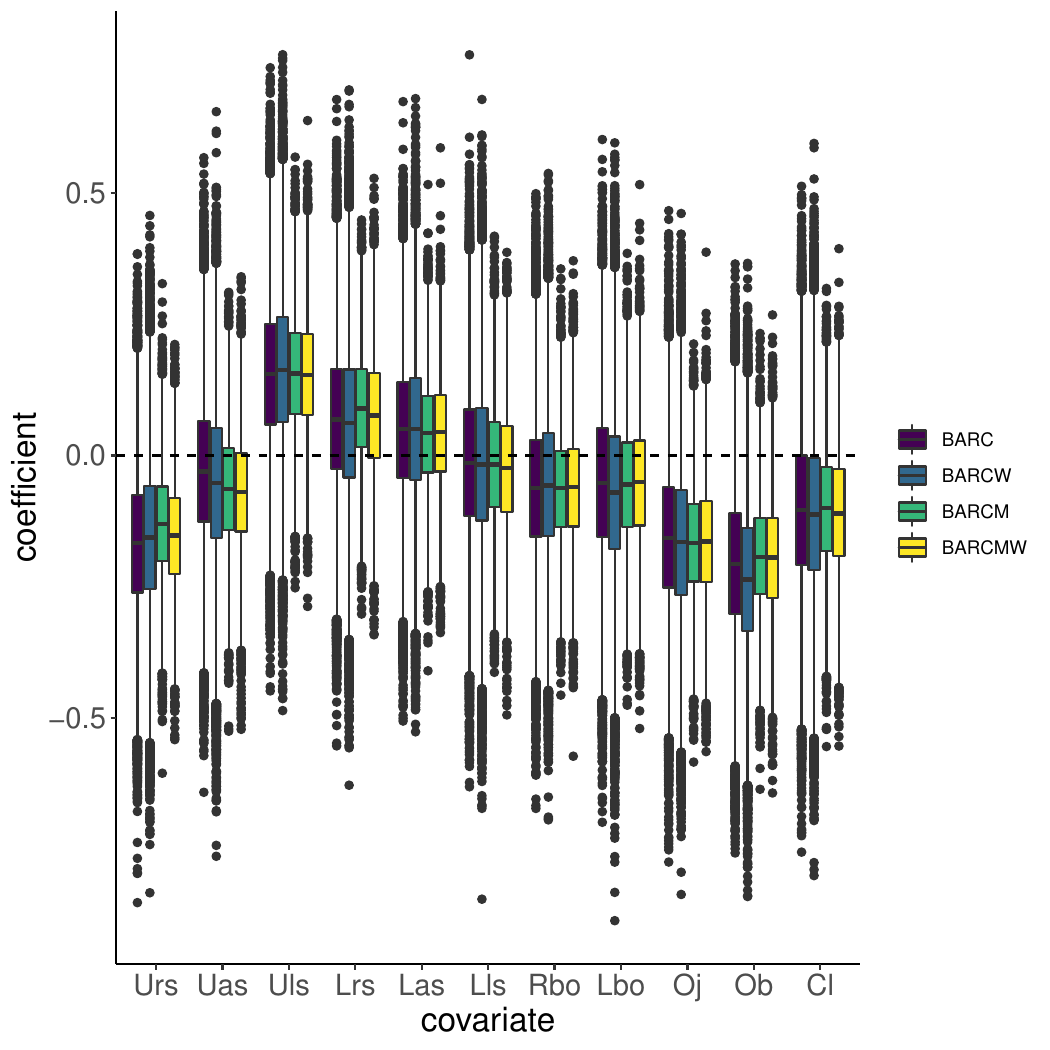}
		\caption{\centering Coefficients under four models} 
	\end{subfigure}%
	\begin{subfigure}{.5\textwidth}
		\centering
		\includegraphics[width=0.9\linewidth]{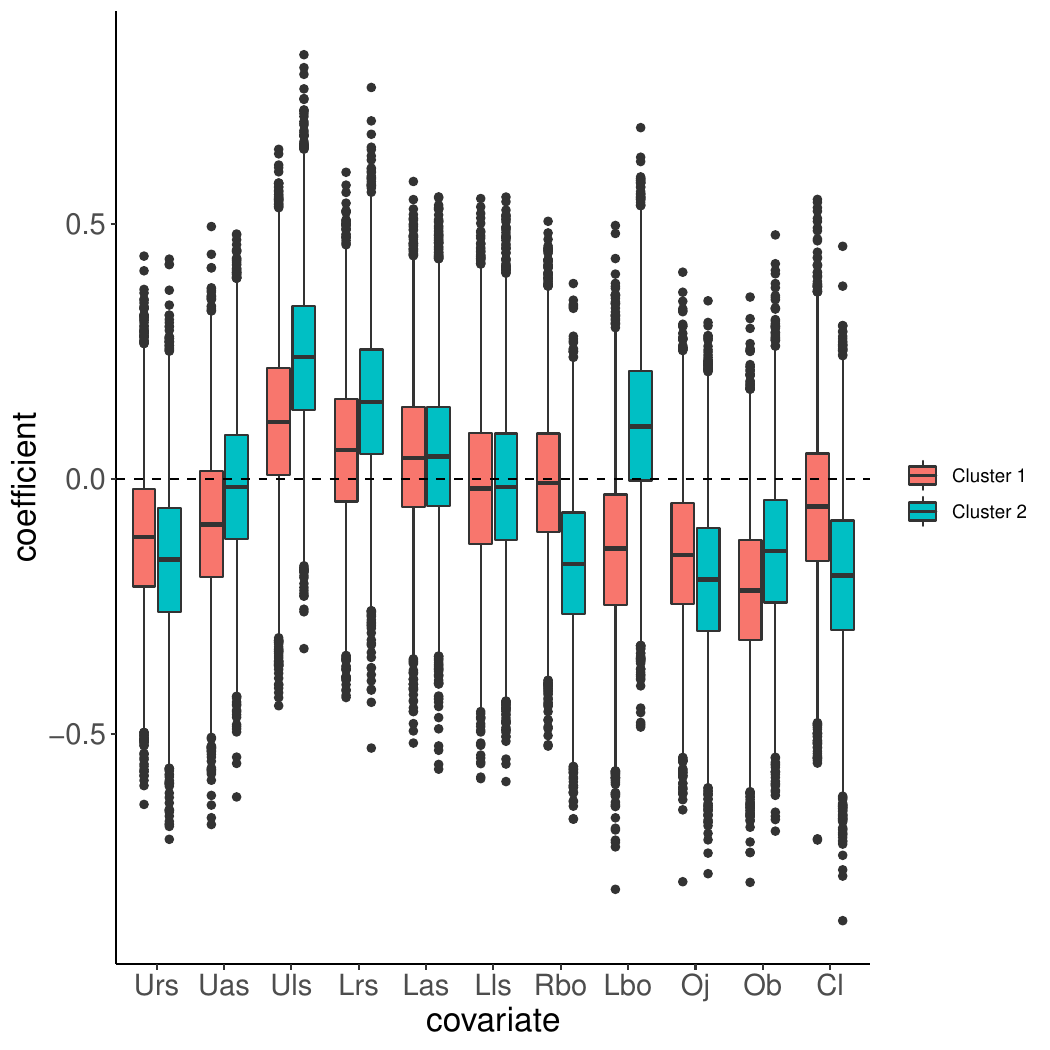}
		\caption{\centering Coefficients for the two clusters from BARCM}
	\end{subfigure}
	\caption{Posterior means and 95$\%$ probability intervals of the coefficients for standardized covariates in orthodontics data under the four Bayesian models and for the two clusters under BARCM. Please refer Table~\ref{tab:ortho_covariate} for the covariate information. 
	}
	\label{fig:Ortho_BARCM_rank_coef}
\end{figure}
Figure \ref{fig:Ortho_BARCM_rank_coef}(a) shows the posterior means and $95\%$ credible intervals of the coefficients $\beta_l$'s under BARC and BARCW,  as well as the average coefficients $m^{-1} \sum_{j=1}^{m} \beta_l^{(j)}$'s under BARCM and BARCMW. 
Under these four models, the covariates have very similar roles in determining the rank, and are crucial 
for positioning patients in those non-overlapping groups. 
In particular, among these 11 covariates, 
overjet, overbite and centerline all measure certain types of overall displacement, and are thus generally considered to have stronger negative effect compared with  the other local displacements in this study. This intuition is further confirmed by our analysis results. 
Figure \ref{fig:Ortho_BARCM_rank_coef}(b) shows the posterior means and $95\%$ credible intervals of the coefficients $\beta_l^{\langle k\rangle}$'s for the two clusters under BARCM. 
Units in clusters 1 and 2 differ by putting different signs on the effect of left buccal occlusion.



\section{Discussion}\label{sec6}

Motivated by two examples we encountered in practice, we reviewed existing literature on the statistical inference of the Thurstone family models for ranking data,  which include the celebrated Thurstone-Mosteller-Daniels model, the Plackett-Luce model, and their various extensions for handling more complex structures and situations, such as when some covariates for the ranked entities  are  observed and/or when the ranking lists may be composed of heterogeneous groups (or equivalently, the rankers may be clustered into different opinion subgroups).

In addition, we described three novel model-based Bayesian rank analysis methods (BARC, BARCW, BARCM), which are based on the TMD modeling framework, unified and extended existing models and methods, and proposed  efficient MCMC algorithms for their needed computations.  With the help of covariates, our new methods can accommodate various types of input ranking lists, including highly incomplete ones. Under the assumption of homogeneous ranking opinion, BARCW learns the qualities of rankers from data, and over-weights high-quality ones in rank aggregation. BARCM, on the other hand, investigates the possibility of having heterogeneous opinion groups among the rankers. 
All three methods evaluate the roles of covariates and generate aggregated ranking lists with uncertainty measures. Our simulation studies and real-data applications validate the importance of covariate information and the estimation of rankers' qualities as well as their heterogeneous opinions.

Our extension to the Thurstone model is similar in spirit to \citet{vitelli2017probabilistic}'s extension of the Mallows model, another popular model for rank data, but we additionally consider the incorporation of covariate information. 
Comparing the Thurstone and Mallows models, the former can be more general in modeling the differences among entities. For example,  any miss ordering of two consecutive entities will have the same probability under the Mallows model with Kendall tau distance, but its probability depends on the underlying true score as well as the noise distribution under the Thurstone model. 
However, the Mallows model can be more robust since the aggregated ranking list minimizes certain average distance from all individual ranking lists, regardless of the underlying data generating process. 
To make the Thurstone model more robust to significant heterogeneity across individual ranking lists, it is of interest to extend the Thurstone model to accommodate more heavy-tailed error distributions, and to develop more efficient MCMC algorithms to deal with the Thurstone mixture models.

In this paper we consider only the covariate information of the ranked entities. It is of interest to further incorporate covariate information of the rankers if such data are available. Rankers' covariates can be helpful for detecting rankers' qualities and clustering rankers into subgroups with different opinions. We leave this extension of BARC and BARCM for a future study. 

\section*{Acknowledgment}
This research is supported in part by the NSF Grants DMS-1712714 and MS-1903139.




\bibliographystyle{apalike} 
\bibliography{rankagg}



\newpage

\setcounter{page}{1}
\begin{center}
	\bf 
	\Large
	\LARGE 
	Supplementary Material for \\
	``Bayesian Analysis of Rank Data with Covariates and Heterogeneous Rankers''
\end{center}

\bigskip

\setcounter{equation}{0}
\setcounter{section}{0}
\setcounter{figure}{0}
\setcounter{example}{0}
\setcounter{proposition}{0}
\setcounter{corollary}{0}
\setcounter{theorem}{0}
\setcounter{table}{0}

\renewcommand {\theproposition} {A\arabic{proposition}}
\renewcommand {\theexample} {A\arabic{example}}
\renewcommand {\thefigure} {A\arabic{figure}}
\renewcommand {\thetable} {A\arabic{table}}
\renewcommand {\theequation} {A\arabic{equation}}
\renewcommand {\thelemma} {A\arabic{lemma}}
\renewcommand {\thesection} {A\arabic{section}}
\renewcommand {\thetheorem} {A\arabic{theorem}}
\renewcommand {\thecorollary} {A\arabic{corollary}}
\renewcommand {\theassumption} {A\arabic{assumption}}

\section{Validity of the parameter-expanded Gibbs sampler}
Below we show the validity of parameter expanded Gibbs sampler under BARC, and the validity under BARCW and BARCM follows by the same logic. We use $\pi$ to denote the marginal posterior distribution of $\bm{Z}$ given all the observed ranking lists $\Tau$, i.e., 
$$
\pi(\bm{Z}) = p(\bm{Z} \mid \Tau) \propto p(\bm{Z}) p(\Tau \mid \bm{Z})
= p(\bm{Z}) 1\{ \text{rank}(\bm{Z}) = \Tau \}.
$$
In order to show the validity of parameter expansion, 
it suffices to prove that for any $\bm{Z}$ following the marginal posterior distribution $\pi(\bm{Z})$, its transformation $t_{\theta}(\bm{Z})$ also follows the same distribution $\pi$, as long as $\theta$ is draw from the distribution with density proportional to $\pi(t_{\theta}(\bm{Z})) 
|J_{\theta}(\bm{Z})| \theta^{-1}$. The proof is as follows.

By construction, the joint density of $(\bm{Z},\theta)$ is 
\begin{align*}
p(\bm{Z},\theta) & = p(\bm{Z}) p(\theta \mid \bm{Z}) = 
\pi(\bm{Z})\cdot 
\frac{
	\pi(t_{\theta}(\bm{Z})) 
	\theta^{-nm-1}}
{
	\int_{\mathbb{R}}
	\pi(t_{\gamma}(\bm{Z})) 
	\gamma^{-nm-1} \text{d}\gamma
},
\end{align*}
which immediately implies the joint density of $(\bm{Y},\theta) \equiv(t_{\theta}(\bm{Z}),\theta)$: 
\begin{align}\label{eq:para_ex_proof}
p(\bm{Y}, \theta) & 
= p(\bm{Z}, \theta) 
|J_{\theta}(\bm{Z})|^{-1}
= 
\pi(\bm{Z} )\cdot 
\frac{
	\pi(t_{\theta}(\bm{Z})) 
	\theta^{-1}}
{
	\int_{\mathbb{R}}
	\pi(t_{\gamma}(\bm{Z})) 
	\gamma^{-nm-1} \text{d}\gamma
}
\nonumber
\\
& = \pi( t_{\theta}^{-1} (\bm{Y}) )\cdot 
\frac{
	\pi(\bm{Y}) 
	\theta^{-1}}
{
	\int_{\mathbb{R}}
	\pi(t_{\gamma}(t_{\theta}^{-1}(\bm{Y})))
	\gamma^{-nm-1} \text{d}\gamma
}.
\end{align}
Note that 
$t_{\gamma}(t_{\theta}^{-1}(\bm{Y})) = \theta \bm{Y}/\gamma = t_{\kappa}^{-1} (\bm{Y})$, where $\kappa = \theta/\gamma$. We can then simplify the denominator in \eqref{eq:para_ex_proof} as 
\begin{align*}
\int_{\mathbb{R}}
\pi(t_{\gamma}(t_{\theta}^{-1}(\bm{Y})))
\gamma^{-nm-1} \text{d}\gamma
&= 
\int_{\mathbb{R}}
\pi(
t_{\kappa}^{-1}(\bm{Y})
)
(\theta/\kappa)^{-nm-1} \text{d}(\theta/\kappa)\\
& = \theta^{-nm} 
\int_{\mathbb{R}}
\pi(
t_{\kappa}^{-1} (\bm{Y})
)
\cdot \kappa^{nm-1} \text{d} \kappa,
\end{align*}
and thus 
further simplify $p(\bm{Y}, \theta)$ as
\begin{align*}
p(\bm{Y}, \theta) 
& = 
\pi(\bm{Y}) \cdot 
\frac{
	\pi( t_{\theta}^{-1} (\bm{Y}) )
	\theta^{-1}}
{
	\theta^{-nm} 
	\int_{\mathbb{R}}
	\pi(
	t_{\kappa}^{-1} (\bm{Y})
	)
	\cdot \kappa^{nm-1} \text{d} \kappa
} = 
\pi(\bm{Y}) \cdot 
\frac{
	\pi( t_{\theta}^{-1} (\bm{Y}) )
	\theta^{nm-1}}
{
	\int_{\mathbb{R}}
	\pi(
	t_{\kappa}^{-1} (\bm{Y})
	)
	\cdot \kappa^{nm-1} \text{d} \kappa
}. 
\end{align*}
Therefore, the marginal density of $\bm{Y}$ is 
\begin{align*}
p(\bm{Y}) & = 
\pi(\bm{Y}) \cdot 
\frac{
	\int_{\mathbb{R}}
	\pi( t_{\theta}^{-1} (\bm{Y}) )
	\theta^{nm-1}
	\text{d} \theta
}
{
	\int_{\mathbb{R}}
	\pi(
	t_{\kappa}^{-1} (\bm{Y})
	)
	\cdot \kappa^{nm-1} \text{d} \kappa
}
= \pi(\bm{Y}),
\end{align*}
i.e., $\bm{Y} \equiv t_{\theta}(\bm{Z})$ follows the distribution with density $\pi$.

\section{Details of Gibbs sampler for the Bayesian models}

\subsection{Gibbs sampler for BARC}
The Gibbs sampler with parameter expansion for BARC model is accomplished by iterating the following steps.
\begin{enumerate}[label=(\arabic*)]
	\item For $j=1,\ldots,M$ and $i=1,\ldots,N$, draw [$Z_{ij} \mid \bm{Z}_{-i,j},\bm{Z}_{-j},\bm{\alpha},\bm{\beta}$] from truncated  $\mathcal{N}(\alpha_i+\bm{x}_{i}^\top \bm{\beta}, \ 1)$, where the truncation points are determined by $\bm{Z}_{-i,j}$ and $\tau_j$
	such that $\text{rank}(\bm{Z}_j)\simeq\tau_j$. 
	
	\item Draw $\theta \sim ( S/\chi^2_{NM} )^{1/2}$ and then update $\bm{Z}$ to be $\bm{Z}/\theta$,  
	where 
	$$
	S=\sum_{j=1}^{M}\bm{Z}_j^\top \bm{Z}_j-\sum_{j=1}^{M}\sum_{j'=1}^{M}\bm{Z}_j^\top \bm{V}(\bm{\Lambda}^{-1}+M\bm{V}^\top \bm{V})^{-1}\bm{V}^\top\bm{Z}_{j'}. 
	$$

	\item	
	Draw $(\bm{\alpha},\bm{\beta}) \sim  \mathcal{N}( \bm{\eta}, \bm{\Sigma} )$, 
	where
	$$
	\bm{\eta}=
	\bm{\Sigma}
	\bm{V}^\top \sum_{j=1}^M \bm{Z}_j
	\ \ 
	\text{  and  } 
	\ \ 
	\bm{\Sigma} = (\bm{\Lambda}^{-1}+M\bm{V}^\top \bm{V})^{-1}.
	$$
	
	\item 
	Draw $\sigma^2_{\alpha} \sim (\nu_\alpha \tau_\alpha^2 + \sum_{i=1}^N \alpha_i^2)/\chi^2_{N+\nu_{\alpha}}$ 
	and $\sigma^2_{\beta} \sim (\nu_\beta \tau_\beta^2 + \sum_{l=1}^L \beta_l^2)/\chi^2_{L+\nu_{\beta}}$. 
\end{enumerate}

\subsection{Gibbs sampler for BARCW}

The Gibbs sampler with parameter expansion for BARCW model is accomplished by iterating the following steps.
\begin{enumerate}[label=(\arabic*)]
	\item For $i=1,\ldots,N$ and $j=1,\ldots,M$, draw [$Z_{ij} \mid \bm{Z}_{-i,j},\bm{Z}_{-j},\bm{\alpha},\bm{\beta}$] from truncated  $\mathcal{N}(\alpha_i+\bm{x}_{i}^\top \bm{\beta},\  w_j^{-1})$ 
	where the truncation points are determined by $\bm{Z}_{-i,j}$ and $\tau_j$	such that $\text{rank}(\bm{Z}_j)\simeq\tau_j$. 
	\item Draw $\theta \sim S^{1/2}/\chi_{NM}$ and then update $\bm{Z}$ to be $\bm{Z}/\theta$,
	where 
	$$
	S=\sum_{j=1}^{M}w_j\bm{Z}_j^\top \bm{Z}_j - 
	\sum_{j=1}^M \sum_{j'=1}^M 
	w_j w_{j'} \bm{Z}_j^\top \bm{V}\left(\bm{\Lambda}^{-1}+\sum_{m=1}^M w_m \bm{V}^\top \bm{V}\right)^{-1}\bm{V}^\top \bm{Z}_{j'}.
	$$ 
	
	\item	
	Draw $(\bm{\alpha},\bm{\beta}) \sim \mathcal{N}( \bm{\eta}, \bm{\Sigma})$, where 
	$$
	\bm{\eta} = 
	\bm{\Sigma}
	\bm{V}^\top \sum_{j=1}^m w_j\bm{Z}_j, 
	\quad 
	\bm{\Sigma} =\left(\bm{\Lambda}^{-1}+\sum_{j=1}^M w_j\bm{V}^\top \bm{V}\right)^{-1}. 
	$$
	
	\item For $j=1,\ldots,M$, 
	draw $w_j$ from a probability mass function proportional to 
	$
	 w_j^{N\over 2}
    e^{-w_j
    \left\| \bm{Z}_j - \bm{\alpha} - \bm{X} \bm{\beta} \right\|_2^2/2}. 
	$
	
	\item Draw $\sigma^2_{\alpha} \sim (\nu_\alpha \tau_\alpha^2 + \sum_{i=1}^N \alpha_i^2)/\chi^2_{N+\nu_{\alpha}}$ 
	and $\sigma^2_{\beta} \sim (\nu_\beta \tau_\beta^2 + \sum_{l=1}^L \beta_l^2)/\chi^2_{L+\nu_{\beta}}$. 
\end{enumerate}

\subsection{Detailed step 2 in Gibbs sampling of BARCM}
The Gibbs sampler with parameter expansion for BARCM model is accomplished by iterating the following steps.
\begin{enumerate}[label=(\arabic*)]
	\item 
	For each $k\in \{c_1,\ldots,c_M\}$, 
	draw $\theta_k \sim S_k^{1/2}/\chi_{N\cdot |\mathcal{R}_k(\bm{c})|}$ and 
	then update $\bm{Z}_j$ to be $\bm{Z}_j/\theta_k$ for $j$ with $c_j = k$, 
	where  
	$$
	S_k =\sum_{j\in \mathcal{R}_k(\bm{c})}\bm{Z}_j^\top \bm{Z}_j - 
	\sum_{j\in \mathcal{R}_k(\bm{c})} 
	\sum_{j'\in \mathcal{R}_k(\bm{c})} 
	\bm{Z}_j^\top \bm{V}\left(\bm{\Lambda}^{-1}+|\mathcal{R}_k(\bm{q})|\bm{V}^\top \bm{V}\right)^{-1}\bm{V}^\top \bm{Z}_{j'}.
	$$ 
	
	\item For each $k\in \{c_1,\ldots,c_M\}$, draw $(\bm{\alpha}^{\langle k\rangle},\bm{\beta}^{\langle k\rangle}) \sim \mathcal{N}\left( \bm{\eta}_k, \bm{\Sigma}_k\right)$, where 
	\begin{align*}
	\bm{\eta}_k & = 
	\bm{\Sigma}_k \bm{V}^\top \sum_{j\in \mathcal{R}_k(\bm{c})}\bm{Z}_j, 
	\ \ 
	\text{ and }
	\ \ 
	\bm{\Sigma}_k  =\left(\bm{\Lambda}^{-1}+|\mathcal{R}_k(\bm{c})|\bm{V}^\top \bm{V}\right)^{-1}.
	\end{align*}
	
	\item For $i=1,\ldots,N$ and $j=1,\ldots,M$, 
	draw [$Z_{ij} \mid \bm{Z}_{-i,j},\bm{Z}_{-j},\bm{\alpha}^{\langle c_j\rangle}, \bm{\beta}^{\langle c_j\rangle}$] from  
	truncated $\mathcal{N}(\alpha^{\langle c_j\rangle}_i+\bm{x}_{i}^\top\bm{\beta}^{\langle c_j\rangle},\ 1)$, 
	where the truncation points are determined by $\bm{Z}_{-i,j}$ and $\tau_j$	such that $\text{rank}(\bm{Z}_j)\simeq\tau_j$. 
    
    \item 
    Let $\mathcal{K} = \{c_1, c_2, \ldots, c_M\}$ be the set of cluster labels of all units, where $\mathcal{K}$ does not contain replicable elements, 
    and $K =|\mathcal{K}|$ be the cardinality of the set $\mathcal{K}$. 
    Draw $\sigma^2_{\alpha} \sim (\nu_\alpha \tau_\alpha^2 + \sum_{k \in \mathcal{K}} \| \bm{\alpha}^{\langle k\rangle} \|_2^2)/\chi^2_{KN+\nu_{\alpha}}$, 
    and 
    $\sigma^2_{\beta} \sim (\nu_\beta \tau_\beta^2 + 
	\sum_{k \in \mathcal{K}} \| \bm{\beta}^{\langle k\rangle} \|_2^2)/\chi^2_{KL+\nu_{\beta}}$. Then draw 
	$\xi \sim \text{Beta}(\gamma+1, n)$, 
	and 
	$\gamma$ from a mixture Gamma distribution 
	$$
	\pi_{\xi} \text{Gamma}(a_{\gamma} + K, b - \log(\xi))
	+ (1-\pi_{\xi}) \text{Gamma}(a_{\gamma} + K-1, b - \log(\xi)),
	$$
	where the weight is defined as 
	$
	\pi_{\xi}/(1-\pi_{\xi}) = (a_{\gamma} + K-1)/\{N(b_{\gamma} -\log(\xi))\}. 
	$
	
	\item 
	For $j=1,\dots,M$, draw $c_j$ from 
	\begin{align*}
		& \quad \ P\left(c_j = k\mid \bm{Z}, \bm{q}_{[-j]}, \Tau\right) \\
		&\propto P\left(c_j = k\mid \bm{c}_{-j}\right)\int p\left(\bm{Z}_{j}\mid \bm{\alpha}^{\langle k\rangle}, \bm{\beta}^{\langle k\rangle}\right)p\left(\bm{\alpha}^{\langle k\rangle}, \bm{\beta}^{\langle k\rangle} \mid \bm{Z}_{-j}\right) \text{d}\bm{\alpha}^{\langle k\rangle}\text{d}\bm{\beta}^{\langle k\rangle}\\
        &\propto P\left(c_j = k\mid \bm{c}_{-j}\right)\cdot\exp\left\{ -\frac{1}{2}h\big( 
        \{j\}\cup \mathcal{R}_k(\bm{c}_{-j})
        \big)+
        \frac{1}{2}
        h\big( 
        \mathcal{R}_k(\bm{c}_{-j})
        \big)
        \right\},
	\end{align*}
	where 
	$P\left(c_j \mid \bm{c}_{-j}\right)$ has the following form: 
	\begin{align*}
		P\left(c_j = k\mid \bm{c}_{-j}\right) & = \frac{|\mathcal{R}_{k}(\bm{q}_{[-j]})|}{(m-1+\gamma)}, 
		\qquad 
		\text{if } 
        k \in \{c_m: m\ne j\}
		\\
		P\left(c_j \notin \{c_m: m\ne j\} \mid \bm{c}_{-j} \right) & = \frac{\gamma}{(m-1+\gamma)},
	\end{align*}
    and $h(\cdot)$ is defined as 
	\begin{align*}
    h(\mathcal{R})
    & = 
    \sum_{m \in \mathcal{R}}
    \bm{Z}_{m}^{\top}\bm{Z}_{m}-
    \sum_{m\in \mathcal{R}} \sum_{m' \in \mathcal{R}} 
    \bm{Z}_{m}^{\top} \bm{V} 
    \left(
    \bm{\Lambda}^{-1}+ \left|\mathcal{R}\right| \bm{V}^{\top} \bm{V} \right)^{-1}
    \bm{V}^\top 
    \bm{Z}_{m'}
    \\
    & \quad \ + \log \left|\bm{\Lambda}^{-1}+ \left| \mathcal{R} \right| \bm{V}^{\top} \bm{V}\right|, 
\end{align*}
with $|\cdot|$ denoting the cardinality of a set or the determinant of a matrix.  
\end{enumerate}

\subsection{Detailed step 2 in Gibbs sampling of BARCMW}\label{gibbs_barcmw}
The Gibbs sampler with parameter expansion for BARCMW model is accomplished by iterating the following steps.
\begin{enumerate}[label=(\arabic*)]
	\item 
	For each $k\in \{c_1,\ldots,c_M\}$, 
	draw $\theta_k \sim S_k^{1/2}/\chi_{N\cdot |\mathcal{R}_k(\bm{c})|}$ and 
	then update $\bm{Z}_j$ to be $\bm{Z}_j/\theta_k$ for $j$ with $c_j = k$, 
	where  
	$$
	S_k =\sum_{j\in \mathcal{R}_k(\bm{c})} w_j \bm{Z}_j^\top \bm{Z}_j - 
	\sum_{j, j'\in \mathcal{R}_k(\bm{c})} w_j w_{j'}
	\bm{Z}_j^\top \bm{V}
	\Big(\bm{\Lambda}^{-1}+ \sum_{m\in \mathcal{R}_k(\bm{c})} w_m \bm{V}^\top \bm{V}\Big)^{-1}
	\bm{V}^\top \bm{Z}_{j'}.
	$$ 
	
	\item For each $k\in \{c_1,\ldots,c_M\}$, draw $(\bm{\alpha}^{\langle k\rangle},\bm{\beta}^{\langle k\rangle}) \sim \mathcal{N}\left( \bm{\eta}_k, \bm{\Sigma}_k\right)$, where 
	\begin{align*}
	\bm{\eta}_k & = 
	\bm{\Sigma}_k \bm{V}^\top \sum_{j\in \mathcal{R}_k(\bm{c})} w_j \bm{Z}_j, 
	\ \ 
	\text{ and }
	\ \ 
	\bm{\Sigma}_k  =\Big(\bm{\Lambda}^{-1}+\sum_{j \in \mathcal{R}_k(\bm{c})} w_j \bm{V}^\top \bm{V}\Big)^{-1}.
	\end{align*}
	
	\item For $i=1,\ldots,N$ and $j=1,\ldots,M$, 
	draw [$Z_{ij} \mid \bm{Z}_{-i,j},\bm{Z}_{-j},\bm{\alpha}^{\langle c_j\rangle}, \bm{\beta}^{\langle c_j\rangle}$] from  
	truncated $\mathcal{N}(\alpha^{\langle c_j\rangle}_i+\bm{x}_{i}^\top\bm{\beta}^{\langle c_j\rangle},\ w_j^{-1})$, 
	where the truncation points are determined by $\bm{Z}_{-i,j}$ and $\tau_j$	such that $\text{rank}(\bm{Z}_j)\simeq\tau_j$. 

    \item For $j=1,\ldots,M$, 
	draw $w_j$ from a probability mass function proportional to 
	$
	 w_j^{N\over 2}
    e^{-w_j
    \left\| \bm{Z}_j - \bm{\alpha}^{\langle c_j\rangle} - \bm{X} \bm{\beta}^{\langle c_j \rangle} \right\|_2^2/2}. 
	$

    \item 
    Let $\mathcal{K} = \{c_1, c_2, \ldots, c_M\}$ be the set of cluster labels of all units, where $\mathcal{K}$ does not contain replicable elements, 
    and $K =|\mathcal{K}|$ be the cardinality of the set $\mathcal{K}$. 
    Draw $\sigma^2_{\alpha} \sim (\nu_\alpha \tau_\alpha^2 + \sum_{k \in \mathcal{K}} \| \bm{\alpha}^{\langle k\rangle} \|_2^2)/\chi^2_{KN+\nu_{\alpha}}$, 
    and 
    $\sigma^2_{\beta} \sim (\nu_\beta \tau_\beta^2 + 
	\sum_{k \in \mathcal{K}} \| \bm{\beta}^{\langle k\rangle} \|_2^2)/\chi^2_{KL+\nu_{\beta}}$. Then draw 
	$\xi \sim \text{Beta}(\gamma+1, n)$, 
	and 
	$\gamma$ from a mixture Gamma distribution 
	$$
	\pi_{\xi} \text{Gamma}(a_{\gamma} + K, b - \log(\xi))
	+ (1-\pi_{\xi}) \text{Gamma}(a_{\gamma} + K-1, b - \log(\xi)),
	$$
	where the weight is defined as 
	$
	\pi_{\xi}/(1-\pi_{\xi}) = (a_{\gamma} + K-1)/\{N(b_{\gamma} -\log(\xi))\}. 
	$
	
	\item 
	For $j=1,\dots,M$, draw $c_j$ from 
	\begin{align*}
		& \quad \ P\left(c_j = k\mid \bm{Z}, \bm{q}_{[-j]}, \Tau\right) \\
		&\propto P\left(c_j = k\mid \bm{c}_{-j}\right)\int p\left(\bm{Z}_{j}\mid \bm{\alpha}^{\langle k\rangle}, \bm{\beta}^{\langle k\rangle}\right)p\left(\bm{\alpha}^{\langle k\rangle}, \bm{\beta}^{\langle k\rangle} \mid \bm{Z}_{-j}\right) \text{d}\bm{\alpha}^{\langle k\rangle}\text{d}\bm{\beta}^{\langle k\rangle}\\
        &\propto P\left(c_j = k\mid \bm{c}_{-j}\right)\cdot\exp\left\{ -\frac{1}{2}h\big( 
        \{j\}\cup \mathcal{R}_k(\bm{c}_{-j})
        \big)+
        \frac{1}{2}
        h\big( 
        \mathcal{R}_k(\bm{c}_{-j})
        \big)
        \right\},
	\end{align*}
	where 
	$P\left(c_j \mid \bm{c}_{-j}\right)$ has the following form: 
	\begin{align*}
		P\left(c_j = k\mid \bm{c}_{-j}\right) & = \frac{|\mathcal{R}_{k}(\bm{c}_{-j})|}{(m-1+\gamma)}, 
		\qquad 
		\text{if } 
        k \in \{c_m: m\ne j\}
		\\
		P\left(c_j \notin \{c_m: m\ne j\} \mid \bm{c}_{-j} \right) & = \frac{\gamma}{(m-1+\gamma)},
	\end{align*}
    and $h(\cdot)$ is defined as 
	\begin{align*}
    h(\mathcal{R})
    & = 
    \sum_{m \in \mathcal{R}} w_m
    \bm{Z}_{m}^{\top}\bm{Z}_{m}-
    \sum_{m\in \mathcal{R}} \sum_{m' \in \mathcal{R}} 
    w_m w_{m'}
    \bm{Z}_{m}^{\top} \bm{V} 
    \left(
    \bm{\Lambda}^{-1}+ \sum_{j' \in \mathcal{R}} w_{j'}  \bm{V}^{\top} \bm{V} \right)^{-1}
    \bm{V}^\top 
    \bm{Z}_{m'}
    \\
    & \quad \ + \log \left|\bm{\Lambda}^{-1}+ \sum_{m \in \mathcal{R}} w_m \bm{V}^{\top} \bm{V}\right|, 
\end{align*}
with $|\cdot|$ denoting the cardinality of a set or the determinant of a matrix.  
\end{enumerate}

\subsection{An R package for the proposed Bayesian models}
We provide an R package \textsf{BayesRankAnalysis} for implementing the proposed Bayesian models for rank data. 
A detailed description for installation and usage of the package can be found on the website \url{https://github.com/li-xinran/BayesRankAnalysis}.

\section{Rank Aggregation Methods in Comparison}\label{methods_compare}
\subsection{Methods based on summary statistics}
Rank aggregation methods based on summary statistics (e.g. average ranking position) are easily understood and widely used. Suppose we have $m$ full ranking lists. Let $\{\tau_j(i)\}_{1\leq j\leq m}$ be the ranking positions of entity $i$ received from all $m$ rankers. The Borda Count method aggregates ranks based on their arithmetic mean, $\sum_{j=1}^{m}\tau_j(i)/m$. 

\subsection{Markov Chain Based Methods}
\cite{dwork2001rank} proposed three Markov Chain based methods ($\text{MC}_1$, $\text{MC}_2$, $\text{MC}_3$) to solve the rank aggregation problem. The basic idea behind these methods is to construct a Markov chain with transition matrix $P=\{p_{i_1i_2}\}_{i_1,i_2\in U}$, where $p_{i_1i_2}$ is the transition probability from entity $i_1$ to entity $i_2$, based on the pairwise comparison information from $\{\tau_1,\ldots,\tau_m\}$. For example, the transition rule of $\text{MC}_2$ is:
\begin{enumerate}
	\item[] If the current state is $i_1$ then the next state is chosen by first picking a list $\tau$ uniformly from all the partial lists $\{\tau_1,\ldots,\tau_m\}$ containing entity $i_1$ then picking an entity $i_2$ uniformly from the set $\{i_2\mid \tau(i_2) \leq \tau(i_1)\}$. 
\end{enumerate}

Then, the authors use the stationary distribution of this Markov chain to generate the aggregated ranking list $\rho$. Explicitly,
$$\rho=\text{sort}(i \in U \text{ by } \pi_i \downarrow),$$
where $\pi=(\pi_1,\ldots,\pi_{|U|})$ satisfies $\pi P=\pi$, and the symbol "$\downarrow$" means that the entities are sorted in descending order.

\subsection{Plackett-Luce based method}
PL model assumes that a ranking list $\tau=[i_1\succ i_2\succ \ldots \succ i_{n}]$ is observed with probability 
$$P(\tau\mid \bm{\gamma}) = \frac{\gamma_{i_1}}{\sum_{l=1}^n\gamma_{i_l}}\times \frac{\gamma_{i_2}}{\sum_{l=2}^n\gamma_{i_l}}\times\cdots\times\frac{\gamma_{i_1}}{\gamma_{i_{n-1}}+\gamma_{i_n}},$$
where $\gamma_{i}\in (0,1)$ and $\sum_{i=1}^n\gamma_i = 1$. Each ranking list from $\{\tau_1,\ldots,\tau_m\}$ follows the above distribution independently. We apply the classical Minorize-Maximization (MM) algorithm for PL model estimation \citep{hunter2004mm}. 

\subsection{Stochastic optimization-based rank aggregation}

Optimization-based rank aggregation methods are proposed to minimize the average distance between a candidate list and each of the input lists, i.e., 
\begin{equation}
\rho=\arg\min_{\sigma\in \mathcal{S}(U)} d(\sigma,\tau_1, \ldots, \tau_m)
\label{eq8}
\end{equation}
where $\mathcal{S}(U)$ represents all allowable rankings, and $d(\cdot)$ is either the average Kendall tau distance or the average Spearman's footrule distance. 
\cite{lin2009integration} used a stochastic search method to optimize \eqref{eq8} by adopting the cross entropy Monte Carlo (CEMC) approach \citep{rubinstein2004cross}. 
In the paper we use the CEMC approach based on the Kendall tau distance.

\end{document}